%% file: 20_10_06_CF-mMIMO_journal.tex
\documentclass[12pt,draftclsnofoot,onecolumn]{IEEEtran}

\IEEEoverridecommandlockouts

\usepackage{algorithm,algorithmic,amsmath,amssymb,amsthm,bbm,cite,color,comment,graphicx,microtype,setspace,url,caption}
\usepackage[USenglish]{babel}
\usepackage[utf8]{inputenc}
\usepackage[T1]{fontenc}

\usepackage{enumitem}
\setitemize{itemsep=0mm,topsep=0mm,parsep=0pt,partopsep=0pt}

\makeatletter
\newcommand\subparagraph{%
  \@startsection{subparagraph}{5}
  {\parindent}
  {3.25ex \@plus 1ex \@minus .2ex}
  {-1em}
  {\normalfont\normalsize\bfseries}}
\makeatother
\usepackage{titlesec}
\let\subparagraph\relax

\usepackage{titlesec}
\titlespacing{\section}{3pt}{4pt plus 2pt minus 1pt}{3pt plus 2pt minus 1pt}
\titlespacing{\subsection}{3pt}{3pt plus 1pt minus 0pt}{2pt plus 1pt minus 0pt}
\usepackage[font=small]{caption}

{}
{}
{}
{}
{\newtheorem{remark}{Remark}}
{}
{}

\input{./Definitions.tex}

\newcommand{\bs}{\textnormal{\tiny{BS}}}
\newcommand{\dl}{\textnormal{\tiny{DL}}}
\newcommand{\mse}{\mathrm{MSE}}
\newcommand{\sinr}{\mathrm{SINR}}
\newcommand{\ue}{\textnormal{\tiny{UE}}}
\newcommand{\ul}{\textnormal{\tiny{UL}}}
\newcommand{\ulA}{\textnormal{\tiny{UL-1}}}
\newcommand{\ulB}{\textnormal{\tiny{UL-2}}}

\title{Distributed Precoding Design via Over-the-Air \\ Signaling for Cell-Free Massive MIMO}
\author{Italo Atzeni,~\IEEEmembership{Member,~IEEE}, Bikshapathi Gouda, \\ and Antti Tölli,~\IEEEmembership{Senior Member,~IEEE}
\thanks{The authors are with the Centre for Wireless Communications, University of Oulu, Finland (emails: \{italo.atzeni, bikshapathi.gouda, antti.tolli\}@oulu.fi). The authors are listed in alphabetical order and the first two authors have contributed~equally to this work.}
\thanks{The work of I.~Atzeni was supported by the Marie Sklodowska-Curie Actions (MSCA-IF 897938 DELIGHT). The work of B.~Gouda and A.~Tölli was supported by the Academy of Finland under grant no. 318927 (6Genesis Flagship).}
\thanks{Part of this work has been presented at the IEEE Int. Workshop Signal Process. Adv. in Wireless Commun. (SPAWC), Atlanta, GA, USA, May 2020~\cite{Gou20}.}}

\begin{document}

\maketitle

\vspace{-10mm}

\begin{abstract}
Most works on cell-free massive multiple-input multiple-output (MIMO) consider non-cooperative precoding strategies at the base stations (BSs) to avoid extensive channel state information (CSI) exchange via backhaul signaling. However, considerable performance gains can be accomplished by allowing coordination among the BSs. This paper proposes the first distributed framework for cooperative precoding design in cell-free massive MIMO (and, more generally, in joint transmission coordinated multi-point) systems that entirely eliminates the need for backhaul signaling for CSI exchange. A novel over-the-air (OTA) signaling mechanism is introduced such that each BS can obtain the same cross-term information that is traditionally exchanged among the BSs via backhaul signaling. The proposed distributed precoding design enjoys desirable flexibility and scalability properties, as the amount of OTA signaling does not scale with the number of BSs or user equipments. Numerical results show fast convergence and remarkable performance gains as compared with non-cooperative precoding design. The proposed scheme can also outperform the centralized precoding design under realistic CSI acquisition.

\end{abstract}

\section{Introduction} \label{sec:INTRO}

Massive multiple-input multiple-output (MIMO) and joint transmission coordinated multi-point (JT-CoMP) are two of the physical-layer wireless technologies that have attracted the most attention during the past ten years. In massive MIMO networks, each base station (BS) is equipped with a large number of antenna elements and serves a smaller number of user equipments (UEs) simultaneously by means of highly directional beamforming techniques~\cite{Mar10,Jun14,Bjo18}. On the other hand, JT-CoMP enables coherent transmission from clusters of cooperating BSs to overcome the inter-cell interference within each cluster~\cite{Jun14,Zha09,Ges10}. While the upcoming 3GPP New Radio (NR) standard for 5G will have massive MIMO as one of its cornerstones~\cite{Eri18}, it will not include JT-CoMP (at least in its first releases) as its implementation in the Long-Term Evolution-Advanced (LTE-A) standard~\cite{Lee12} did not achieve significant gains in practice. This can be mainly attributed to the considerable amount of backhaul signaling required for channel state information (CSI) and data sharing~\cite{Bas17} but also to a network-centric approach to coherent transmission~\cite{Int19}, whereby the BSs in a cluster cooperate to serve the UEs in their joint coverage region. The practical implementation of JT-CoMP was also hindered by other attributes of LTE-A, such as a frequency division duplex dominated macro-cell deployment and a rigid frame/slot structure in its time division duplex (TDD) mode of operation, which did not allow for a flexible channel estimation.

Cell-free massive MIMO~\cite{Int19,Zha19} is a recently coined concept that conveniently combines elements from massive MIMO~\cite{Bjo18}, small cells~\cite{Jun14}, and UE-centric JT-CoMP~\cite{Wan16}. In a cell-free context, the massive MIMO regime is achieved by spreading a large number of low-cost access points across the network (even in the form of single-antenna BSs~\cite{Ngo17,Nay17}), which provides enhanced coverage and reduced pathloss. All the BSs operate in TDD mode and jointly serve all the UEs, which ideally allows to entirely eliminate the inter-cell interference. To this end, all the BSs are assumed to be connected to a central processing unit (CPU) by means of backhaul links that provide the UE-specific data and, if required by the adopted physical-layer transmission scheme, enable network-wide processing for the computation of the precoding strategies. More recent works advocate a purely UE-centric approach to coherent transmission, where clusters of cooperating BSs are formed so that each UE is served by its nearest BSs~\cite{Buz20}. Both the ``all serve all'' and the UE-centric views described above represent a sharp departure from traditional JT-CoMP, which is normally implemented in a network-centric fashion with well-defined and non-overlapping clusters of BSs~\cite{Zha09}.

Cell-free massive MIMO has been the subject of an extensive literature over the past few years and is now regarded as a potential physical-layer paradigm shift for beyond-5G systems~\cite{Zha19,Raj20,Raj20a,Zha20}. Remarkably, cell-free massive MIMO networks have been shown to outperform traditional cellular massive MIMO and small-cell networks in several practical scenarios~\cite{Nay17,Ngo17,Bjo20,Liu20,Pap20}. Their performance has been analyzed under several realistic network and hardware assumptions, e.g., with hybrid analog-digital precoding~\cite{Alo19,Fem19}, with low-resolution analog-to-digital converters~\cite{Hu19,Zha19a}, under channel non-reciprocity~\cite{Pal20}, as well as with hardware impairments and limited backhaul capacity~\cite{Zha18,Fem19,Mas20}. Another important focus is the global energy efficiency, which has been studied considering the impact of backhaul power consumption~\cite{Ngo18} and quantization~\cite{Bas19a} among other factors. To avoid CSI exchange among the BSs via backhaul signaling and to reduce the overall computational complexity, most of the aforementioned works (i.e.,~\cite{Int19,Buz20,Ngo17,Zha20,Pap20,Fem19,Hu19,Zha19a,Pal20,Zha18,Mas20,Ngo18,Bas19a}) assume simple non-cooperative precoding strategies at the BSs, such as matched filtering (MF), local zero-forcing (ZF), and local minimum mean squared error (MMSE) precoding, which can be implemented based on locally acquired CSI (see also~\cite{Int20}). However, the performance of cell-free massive MIMO systems can be considerably improved by increasing the level of coordination among the BSs~\cite{Bjo20}.

Cooperative precoding design for JT-CoMP can be broadly classified into centralized and distributed approaches. In the centralized precoding design, the BSs forward their locally acquired CSI to the CPU via backhaul signaling and the CPU feeds back the optimized precoding strategies to the BSs. Here, both the amount of CSI exchange between the BSs and the CPU and the computational complexity of the precoding optimization at the CPU may become overwhelming due to the high dimensionality of the aggregated channels. In the cell-free massive MIMO literature, such a centralized approach has been adopted, e.g., in~\cite{Nay17,Liu20,Alo19}, which assume centralized ZF precoding/combining, and in~\cite{Bjo20}, which considers centralized MMSE combining among different levels of coordination. To avoid the centralized computation,~\cite{Kal18} proposed a distributed iterative framework for JT-CoMP that allows to optimize the precoding strategies locally at each BS using bi-directional training between the BSs and the UEs~\cite{Tol19} in addition to periodic exchange of cross-term information among nearby BSs via backhaul signaling. Despite a significant complexity reduction, the extensive CSI exchange among the BSs makes the practical implementation of~\cite{Kal18} challenging (see also~\cite{Bas17,Zha20}); furthermore, the backhaul introduces delays and quantization errors into the CSI exchange that can sensibly degrade the performance of the precoding design. These issues are particularly critical in a cell-free massive MIMO context due to the large number of BSs and UEs involved in the joint processing.

\subsection{Contribution}

Non-cooperative precoding strategies (such as MF, local ZF, and local MMSE precoding) have been so far preferred in the cell-free massive MIMO literature as they do not require any CSI exchange via backhaul signaling. However, the fact that the channel hardening effect is less pronounced in cell-free massive MIMO than in cellular massive MIMO~\cite{Int19} suggests that cooperative precoding design can bring considerable performance gains over its non-cooperative counterpart. In this paper, we bridge this gap and propose the first distributed framework for cooperative precoding design in cell-free massive MIMO (and, more generally, in JT-CoMP) systems that entirely eliminates the need for backhaul signaling for CSI exchange. Focusing on the weighted sum mean squared error (MSE) minimization, a novel over-the-air (OTA) signaling mechanism allows each BS to obtain the same cross-term information that was exchanged among the BSs via backhaul signaling in~\cite{Kal18}. Specifically, this is achieved by introducing a new uplink signaling resource and a new CSI combining mechanism that complement the existing uplink and downlink pilot-aided channel estimations. The proposed distributed precoding design enjoys desirable flexibility and scalability properties, as the amount of OTA signaling does not scale with the number of BSs or UEs; furthermore, there are no delays in the CSI exchange among the BSs. These practical benefits come at the cost of extra uplink signaling overhead per bi-directional training iteration, which, however, results in a minor performance loss with respect to the distributed precoding design via backhaul signaling.

The contributions of this paper are summarized as follows:
\begin{itemize}[leftmargin=6mm]
\item[$\bullet$] Building on existing tools from JT-CoMP and considering multi-antenna UEs, we describe centralized and distributed precoding schemes for cell-free massive MIMO under both perfect CSI and realistic pilot-aided CSI acquisition.
\item[$\bullet$] We propose a distributed precoding design where the CSI exchange among the BSs via backhaul signaling in~\cite{Kal18} is entirely replaced by a novel OTA signaling mechanism, which does not scale with the number of BSs or UEs.
\item[$\bullet$] We address relevant implementation aspects of the proposed distributed precoding design and illustrate how the OTA signaling can be integrated into the flexible 5G 3GPP NR frame/slot structure~\cite{3GPP_38.211}.
\item[$\bullet$] Numerical results show significant performance gains in terms of average sum rate over non-cooperative precoding design even after a small number of iterations; remarkably, the proposed distributed precoding design via OTA signaling outperforms its centralized counterpart in presence of imperfect CSI and the huge practical benefits with respect to the case with ideal backhaul signaling come at the cost of a very modest performance loss.
\end{itemize}

\textit{Outline.} The rest of the paper is structured as follows. Section~\ref{sec:SM} introduces the cell-free massive MIMO system model. Section~\ref{sec:PROB} describes the centralized and the distributed precoding design with perfect CSI. Then, Section~\ref{sec:PROB_IMP} extends the previous section by considering realistic pilot-aided CSI acquisition. As the main contribution of this paper, Section~\ref{sec:OTA} presents the distributed precoding design via OTA signaling. In Section~\ref{sec:NUM}, numerical results are reported to illustrate the remarkable performance of the proposed scheme in different practical scenarios. Finally, Section~\ref{sec:CONCL} summarizes our contributions and draws some concluding remarks.

\textit{Notation.} Lowercase and uppercase boldface letters denote vectors and matrices, respectively, whereas $(\cdot)^{\tran}$ and $(\cdot)^{\herm}$ are the transpose and Hermitian transpose operators, respectively. $\| \cdot \|$ and $\| \cdot \|_{\rmF}$ represent the Euclidean norm for vectors and the Frobenius norm for matrices, respectively. $\Re[\cdot]$ and $\Exp[\cdot]$ are the real part and expectation operators, respectively. $\I_{L}$ denotes the $L$-dimensional identity matrix and $\0$ represents the zero vector or matrix with proper dimension. $\tr(\cdot)$ is the trace operator and $\Diag(\cdot)$ produces a diagonal matrix with the elements of the vector argument on the diagonal. $[a_{1}, \ldots, a_{L}]$ de$\H_{b,k} \in \Compl^{M \times N}$notes horizontal concatenation, whereas $\{a_{1}, \ldots, a_{L}\}$ or $\{ a_{\ell} \}_{\ell \in \setL}$ denote the set of elements in the argument. Lastly, $\setC \setN(0, \sigma^{2})$ is the complex normal distribution with zero mean and variance $\sigma^{2}$, whereas $\nabla_{\x}(\cdot)$ denotes the gradient with respect~to~$\x$.

\section{System Model} \label{sec:SM}

Consider a downlink cell-free massive MIMO network where a set of BSs $\setB \triangleq \{1, \ldots, B\}$, each equipped with $M$ antennas, serves a set of UEs $\setK \triangleq \{1, \ldots, K\}$, each equipped with $N$ antennas.\footnote{For notational simplicity, we assume the same number of antennas for all the BSs and the UEs. However, the subsequent analysis and algorithms are valid for any number of antennas.} Assuming a TDD setting and, for simplicity, a single data stream per UE, let $\H_{b,k} \in \Compl^{M \times N}$ be the uplink channel matrix between UE~$k \in \setK$ and BS~$b \in \setB$, with $\H_{k} \triangleq [\H_{1,k}^{\tran}, \ldots, \H_{B,k}^{\tran}]^{\tran} \in \Compl^{B M \times N}$ denoting the aggregated uplink channel matrix of UE~$k$. Likewise, let $\w_{b,k} \in \Compl^{M \times 1}$ be the BS-specific precoding vector used by BS~$b$ for UE~$k$, with $\w_{k} \triangleq [\w_{1,k}^{\tran}, \ldots, \w_{B,k}^{\tran}]^{\tran} \in \Compl^{B M \times 1}$ denoting the aggregated precoding vector used for UE~$k$; here, we assume the per-BS power constraints $\big\{ \sum_{k \in \setK} \|\w_{b,k}\|^{2} \leq \rho_{\bs} \big\}_{b \in \setB}$, where $\rho_{\bs}$ denotes the maximum transmit power at each BS.\footnote{Here, $\rho_{\bs}$ does not include the hardware power consumption. The impact of this factor is considered, e.g., in~\cite{Ngo18} in the context of power control and in~\cite{Fan20} in the context of hybrid analog-digital precoding.} Note that, according to the previous definitions, we have $\H_{k}^{\herm} \w_{\bar{k}} = \sum_{b \in \setB} \H_{b,k}^{\herm} \w_{b,\bar{k}}$. Hence, the receive signal at UE~$k$ is given by
\begin{align} \label{eq:y_k}
\y_{k} & \triangleq \sum_{b \in \setB} \sum_{\bar{k} \in \setK} \H_{b,k}^{\herm} \w_{b,\bar{k}} d_{\bar{k}} + \z_{k} \in \Compl^{N \times 1}
\end{align}
where $d_{k} \sim \setC \setN (0,1)$ is the transmit data symbol for UE~$k$ and $\z_{k}$ is the average white Gaussian noise (AWGN) term at UE~$k$ with elements distributed as $\setC \setN (0, \sigma_{k}^{2})$. Upon receiving $\y_{k}$, UE~$k$ uses the combining vector $\v_{k} \in \Compl^{N \times 1}$ to combine $\y_{k}$ and the resulting signal-to-interference- plus-noise ratio (SINR) reads as
\begin{align} \label{eq:SINR_k}
\sinr_{k} & \triangleq \frac{|\sum_{b \in \setB} \v_{k}^{\herm} \H_{b,k}^{\herm} \w_{b,k}|^{2}}{\sum_{\bar{k} \in \setK \setminus \{k\}} |\sum_{b \in \setB} \v_{k}^{\herm} \H_{b,k}^{\herm} \w_{b,\bar{k}}|^{2} + \sigma_{k}^{2} \| \v_{k} \|^{2}}.
\end{align}
From \eqref{eq:SINR_k}, it is easy to observe that the design of the precoding vectors depends on the combining vectors and vice versa. Finally, the sum rate (measured in bps/Hz) is given by
\begin{align} \label{eq:R}
R \triangleq \sum_{k \in \setK} \log_{2}(1 + \sinr_{k}).
\end{align}
Note that the above sum rate, which uses the SINR expression in \eqref{eq:SINR_k}, represents an upper bound on the system performance for fixed precoding and combining vectors, which is achievable if perfect global CSI is available at all the BSs~\cite{Cai18}. The average sum rate based on \eqref{eq:R} will be considered in Section~\ref{sec:NUM} to evaluate and compare the different precoding schemes.

This paper focuses on distributed precoding design, where each BS~$b$ optimizes its precoding vectors $\{\w_{b,k}\}_{k \in \setK}$ locally while coordinating with the other BSs. For the sake of comparison, we also illustrate the centralized precoding design, where the aggregated precoding vectors $\{\w_{k}\}_{k \in \setK}$ are optimized by the CPU and the BS-specific precoding vectors are fed back to the BSs. In both cases, the combining vectors $\{\v_{k}\}_{k \in \setK}$ are computed locally by the corresponding UEs. In the following, we describe realistic pilot-aided CSI acquisition at both the BSs (in Section~\ref{sec:SM_ul}) and the UEs (in Section~\ref{sec:SM_dl}), which will be heavily referred to in Sections~\ref{sec:PROB_IMP}~and~\ref{sec:OTA} as part of the adopted bi-directional training.

\subsection{Uplink Pilot-Aided Channel Estimation} \label{sec:SM_ul}

Let $\h_{b,k} \triangleq \H_{b,k} \v_{k} \in \Compl^{M \times 1}$ be the effective uplink channel vector between UE~$k$ and BS~$b$, and let $\p_{k} \in \Compl^{\tau \times 1}$ be the pilot sequence assigned to UE~$k$, with $\|\p_{k}\|^{2} = \tau$. Moreover, let $\rho_{\ue}$ denote the maximum transmit power at each UE. In the uplink pilot-aided channel estimation phase, each UE~$k$ synchronously transmits its pilot sequence $\p_{k}$ using its combining vector $\v_{k}$ as precoder, i.e., \vspace{-1mm}
\begin{align} \label{eq:X_k_ul1}
\X_{k}^{\ulA} \triangleq \sqrt{\beta^{\ulA}} \v_{k} \p_{k}^{\herm} \in \Compl^{N \times \tau}
\end{align}
where the power scaling factor $\beta^{\ulA}$ (equal for all the UEs) ensures that $\X_{k}^{\ulA}$ complies with the UE transmit power constraint (see Section~\ref{sec:OTA_power} for more details on the choice of $\beta^{\ulA}$). Then, the receive signal at BS~$b$ is given by

\clearpage

$ $ \vspace{-18mm}

\begin{align}
\Y_{b}^{\ulA} & \triangleq \sum_{k \in \setK} \H_{b,k} \X_{k}^{\ulA} + \Z_{b}^{\ulA} \\
\label{eq:Y_b_ul1} & = \sqrt{\beta^{\ulA}} \sum_{k \in \setK} \h_{b,k} \p_{k}^{\herm} + \Z_{b}^{\ulA} \in \Compl^{M \times \tau}
\end{align}
where $\Z_{b}^{\ulA} \in \Compl^{M \times \tau}$ is the AWGN term at BS~$b$ with elements distributed as $\setC \setN (0, \sigma_{b}^{2})$, and the least-squares (LS) estimate of $\h_{b,k}$ is obtained as
\begin{align}
\hat{\h}_{b,k} & \triangleq \frac{1}{\tau \sqrt{\beta^{\ulA}}} \Y_{b}^{\ulA} \p_{k} \\
\label{eq:h_bk_hat} & = \h_{b,k} + \frac{1}{\tau} \sum_{\bar{k} \in \setK \setminus \{k\}} \h_{b,\bar{k}} \p_{\bar{k}}^{\herm} \p_{k} + \frac{1}{\tau \sqrt{\beta^{\ulA}}} \Z_{b}^{\ulA} \p_{k}.
\end{align}
Here, perfect channel estimation is achieved when:
\begin{itemize}[leftmargin=6mm]
\item[\textit{i)}] The pilot contamination in the second term of \eqref{eq:h_bk_hat} is eliminated using, for instance, orthogonal pilots (i.e., $\{\p_{\bar{k}}^{\herm} \p_{k} = 0\}_{\bar{k} \in \setK \setminus \{k\}}$) or non-orthogonal random pilots with infinite pilot length (i.e., $\tau \to \infty$);
\item[\textit{ii)}] The channel estimation noise in the third term of \eqref{eq:h_bk_hat} is eliminated using infinite pilot length.
\end{itemize}
Note that these observations also apply to \eqref{eq:H_bk_hat} and \eqref{eq:g_k_hat} in the following.

On the other hand, the estimation of the channel matrix $\H_{b,k}$, which is required in the centralized precoding design, involves $N$ antenna-specific pilot sequences for UE~$k$. In this context, let $\P_{k} \in \Compl^{\tau \times N}$ be the pilot matrix assigned to UE~$k$, with $\|\P_{k}\|_{\rmF}^{2} = \tau N$. In the uplink pilot-aided channel estimation phase, each UE~$k$ synchronously transmits its pilot matrix, i.e., \vspace{-2.5mm}
\begin{align} \label{eq:X_k_ul}
\X_{k}^{\ul} \triangleq \sqrt{\beta^{\ul}} \P_{k}^{\herm}
\end{align}

\vspace{-1.5mm}

\noindent where the power scaling factor $\beta^{\ul} \triangleq \frac{\rho_{\ue}}{N}$ ensures that $\X_{k}^{\ul}$ complies with the UE transmit power constraint. Then, the receive signal at BS~$b$ is given by \vspace{-1.5mm}
\begin{align}
\Y_{b}^{\ul} & \triangleq \sum_{k \in \setK} \H_{b,k} \X_{k}^{\ul} + \Z_{b}^{\ul} \\
\label{eq:Y_b_ul} & = \sqrt{\beta^{\ul}} \sum_{k \in \setK} \H_{b,k} \P_{k}^{\herm} + \Z_{b}^{\ul} \in \Compl^{M \times \tau}
\end{align}

\vspace{-1mm}

\noindent where $\Z_{b}^{\ul} \in \Compl^{M \times \tau}$ is the AWGN term at BS~$b$ with elements distributed as $\setC \setN (0, \sigma_{b}^{2})$, and the LS estimate of $\H_{b,k}$ is obtained as \vspace{-1mm}
\begin{align}
\hat{\H}_{b,k} & \triangleq \frac{1}{\tau \sqrt{\beta^{\ul}}} \Y_{b}^{\ul} \P_{k} \\
\label{eq:H_bk_hat} & = \frac{1}{\tau} \sum_{\bar{k} \in \setK} \H_{b,\bar{k}} \P_{\bar{k}}^{\herm} \P_{k} + \frac{1}{\tau \sqrt{\beta^{\ul}}} \Z_{b}^{\ulA} \P_{k} \\
\label{eq:H_bk_hat_orth} & = \H_{b,k} + \frac{1}{\tau} \sum_{\bar{k} \in \setK \setminus \{k\}} \H_{b,\bar{k}} \P_{\bar{k}}^{\herm} \P_{k} + \frac{1}{\tau \sqrt{\beta^{\ul}}} \Z_{b}^{\ul} \P_{k}
\end{align}

\noindent where \eqref{eq:H_bk_hat_orth} holds only if $\P_{k}^{\herm} \P_{k} = \tau \I_{N}$ (i.e., if there is no pilot contamination among the columns of $\P_{k}$).

\vspace{-2mm}
\begin{remark} \rm{
The expressions of the receive signals in \eqref{eq:Y_b_ul1} and \eqref{eq:Y_b_ul} imply that the transmit signals from all the UEs are received synchronously by each BS. Although perfect synchronization is infeasible, quasi-synchronous operations can be achieved in practice by setting the duration of the cyclic prefix to accommodate both the synchronization errors and the delay spread, as described in~\cite{Int19,Bjo20}. These considerations also apply to the downlink pilot-aided channel estimation in Section~\ref{sec:SM_dl} (see \eqref{eq:Y_k_dl}) and to the new uplink signaling resource introduced in Section~\ref{sec:OTA}~(see~\eqref{eq:Y_b_ul2}).}
\end{remark} \vspace{-2mm}

\vspace{-0.5mm}

\subsection{Downlink Pilot-Aided Channel Estimation} \label{sec:SM_dl}

Let $\g_{k} \triangleq \sum_{b \in \setB} \H_{b,k}^{\herm} \w_{b,k} \in \Compl^{N \times 1}$ be the effective downlink channel vector between all the BSs and UE~$k$. In the downlink pilot-aided channel estimation phase, each BS~$b$ synchronously transmits a superposition of the pilot sequences $\{\p_{k}\}_{k \in \setK}$ after precoding them with the corresponding precoding vectors $\{\w_{b,k}\}_{k \in \setK}$, i.e.,
\begin{align} \label{eq:X_b_dl}
\X_{b}^{\dl} \triangleq \sum_{k \in \setK} \w_{b,k} \p_{k}^{\herm} \in \Compl^{M \times \tau}.
\end{align}
Then, the receive signal at UE~$k$ is given by
\begin{align}
\Y_{k}^{\dl} & \triangleq \sum_{b \in \setB} \H_{b,k}^{\herm} \X_{b}^{\dl} + \Z_{k}^{\dl} \\
\label{eq:Y_k_dl} & = \sum_{b \in \setB} \sum_{\bar{k} \in \setK} \H_{b,k}^{\herm} \w_{b,\bar{k}} \p_{\bar{k}}^{\herm} + \Z_{k}^{\dl} \in \Compl^{N \times \tau}
\end{align}
where $\Z_{k}^{\dl} \in \Compl^{N \times \tau}$ is the AWGN term at UE~$k$ with elements distributed as $\setC \setN (0, \sigma_{k}^{2})$, and the LS estimate of $\g_{k}$ is obtained as
\begin{align}
\hat{\g}_{k} & \triangleq \frac{1}{\tau} \Y_{k}^{\dl} \p_{k} \\
\label{eq:g_k_hat} & = \g_{k} + \frac{1}{\tau} \sum_{b \in \setB} \sum_{\bar{k} \in \setK \setminus \{k\}} \H_{b,k}^{\herm} \w_{b,\bar{k}} \p_{\bar{k}}^{\herm} \p_{k} + \frac{1}{\tau} \Z_{k}^{\dl} \p_{k}.
\end{align}
Note that most of the papers on cell-free massive MIMO mentioned in Section~\ref{sec:INTRO} do not consider downlink channel estimation as they assume a simplified system model with single-antenna UEs where there are no transmit/receive strategies to optimize at the latter. On the other hand, in the context of multi-antenna UEs, the downlink channel estimation phase is paramount for the optimization of such strategies. Furthermore, it is shown in~\cite{Int19a} that estimating the downlink channels helps to compensate for the less pronounced channel hardening effect with respect to cellular massive MIMO, even for single-antenna UEs.

\section{Problem Formulation with Perfect CSI} \label{sec:PROB}

In this paper, we target the weighted sum MSE minimization problem to optimize the precoding vectors $\{\w_{b,k}\}_{b \in \setB, k \in \setK}$ and the combining vectors $\{\v_{k}\}_{k \in \setK}$. This can be used as a surrogate of the more involved weighted sum rate maximization problem (or, equivalently, of the iterative weighted sum MSE minimization problem~\cite{Shi11}). In fact, since the total number of BS antennas in the network~$B M$ is much larger than the number of UEs~$K$, the weighted sum MSE minimization yields only a minor penalty in terms of sum-rate performance as compared with the weighted sum rate maximization, while being much easier to handle and providing an inherent fairness across the UEs. In this section, we tackle the weighted sum MSE minimization problem under perfect channel estimation; the results derived here will be highly useful to describe the case of realistic pilot-aided CSI acquisition at both the BSs and the UEs in Sections~\ref{sec:PROB_IMP}~and~\ref{sec:OTA}.

Let $\omega_{k}$ denote the weight assigned to UE~$k$, which is fixed before the transmission (e.g., by the CPU) to capture the UE's priority. Building on \eqref{eq:y_k}, let us introduce the MSE at UE~$k$ as
\begin{align}
\mse_{k} & \triangleq \Exp \big[ |\v_{k}^{\herm} \y_{k} - d_{k}|^{2} \big] \\
\label{eq:MSE_k} & = \sum_{\bar{k} \in \setK} \bigg| \sum_{b \in \setB} \v_{k}^{\herm} \H_{b,k}^{\herm} \w_{b,\bar{k}} \bigg|^{2} - 2 \Re \bigg[ \sum_{b \in \setB} \v_{k}^{\herm} \H_{b,k}^{\herm} \w_{b,k} \bigg] + \sigma_{k}^{2} \| \v_{k} \|^{2} + 1.
\end{align}
The sum MSE, i.e., $\sum_{k \in \setK} \omega_{k} \mse_{k}$, is convex with respect to either the transmit or the receive strategies, but not jointly convex with respect to both. This makes the joint optimization of the precoding and the combining vectors extremely challenging, especially under limited signaling between the BSs and the UEs. Hence, we can achieve a local optimum of the sum MSE minimization problem by using \textit{alternating optimization}, whereby the precoding vectors are optimized for fixed combining vectors and vice versa in an iterative best-response fashion (as done, e.g., in~\cite{Shi11,Kal18}).

\vspace{1mm}

\begin{itemize}[leftmargin=6mm]
\item[$\bullet$] \textbf{Optimization of the combining strategies.} The combining vectors $\{\v_{k}\}_{k \in \setK}$ are computed locally and independently by the UEs such that each UE~$k$ minimizes $\mse_{k}$ in \eqref{eq:MSE_k}. In the centralized precoding design, the combining vectors are also derived by the CPU in conjunction with the precoding vectors as part of the alternating optimization routine (although they are not fed back to the UEs). From the point of view of UE~$k$, we can rewrite~the~MSE~as

\clearpage

$ $ \vspace{-17mm}

\begin{align} \label{eq:MSE_k_v}
\mse_{k} = \v_{k}^{\herm} (\Psib_{k} + \sigma_{k}^{2} \I_{N}) \v_{k} - 2 \Re [\v_{k}^{\herm} \g_{k}] + 1
\end{align}
where we have defined
\begin{align} \label{eq:Psi_k}
\Psib_{k} \triangleq \sum_{\bar{k} \in \setK} \bigg( \sum_{b \in \setB} \H_{b,k}^{\herm} \w_{b,\bar{k}} \bigg) \bigg( \sum_{b \in \setB} \w_{b,\bar{k}}^{\herm} \H_{b,k} \bigg).
\end{align}
The combining vector $\v_{k}$ that minimizes \eqref{eq:MSE_k_v} is the well-known MMSE receiver, which may be written as \vspace{-3mm}
\begin{align} \label{eq:v_k}
\v_{k} = (\Psib_{k} + \sigma_{k}^{2} \I_{N})^{-1} \g_{k}.
\end{align}
Observe that $\v_{k}$ can be computed locally by UE~$k$ as in \eqref{eq:v_k} if $\Psib_{k}$ in \eqref{eq:Psi_k} and the effective downlink channel $\g_{k}$ are known by UE~$k$.

\vspace{1mm}

\item[$\bullet$] \textbf{Optimization of the precoding strategies.} The precoding vectors $\{\w_{b,k}\}_{b \in \setB, k \in \setK}$ are computed as the solutions of the weighted sum MSE minimization problem with per-BS power constraints. To this end, we introduce the following preliminary definitions: $\h_{k} \triangleq [\h_{1,k}^{\tran}, \ldots, \h_{B,k}^{\tran}]^{\tran} \in \Compl^{B M \times 1}$, $\H \triangleq [\h_{1}, \ldots, \h_{K}] \in \Compl^{B M \times K}$, $\W \triangleq [\w_{1}, \ldots, \w_{K}] \in \Compl^{B M \times K}$, $\Omegab \triangleq \Diag \big( [\omega_{1}, \ldots, \omega_{K}] \big) \in \Real^{K \times K}$, and $\Phib \triangleq \sum_{k \in \setK} \omega_{k} \h_{k} \h_{k}^{\herm} \in \Compl^{B M \times B M}$, where the latter may be rewritten as
\begin{align} \label{eq:Phi}
\Phib \triangleq
\begin{bmatrix}
\Phib_{1 1}         & \ldots    & \Phib_{1 B} \\
\vdots              & \ddots    & \vdots \\
\Phib_{1 B}^{\herm} & \ldots    & \Phib_{B B}
\end{bmatrix}
\end{align}

\vspace{1mm}

\noindent with $\Phib_{b \bar{b}} \triangleq \sum_{k \in \setK} \omega_{k} \h_{b,k} \h_{\bar{b},k}^{\herm} \in \Compl^{M \times M}$. Finally, the weighted sum MSE can be expressed as
\begin{align}
\label{eq:sum_MSE} \sum_{k \in \setK} \omega_{k} \mse_{k} = \tr(\W^{\herm} & \Phib \W) - 2 \Re \big[ \tr(\Omegab \H^{\herm} \W) \big] + \sum_{k \in \setK} \omega_{k} (\sigma_{k}^{2} \| \v_{k} \|^{2} + 1).
\end{align}
In the following, we first describe the centralized precoding design in Section~\ref{sec:PROB_centr} and then focus on the distributed precoding design via backhaul signaling in Section~\ref{sec:PROB_distr}.
\end{itemize}

\subsection{Centralized Precoding Design} \label{sec:PROB_centr}

In the centralized precoding design, the aggregated precoding vectors are computed by the CPU and the BS-specific precoding vectors are fed back to the corresponding BSs via backhaul signaling. Here, the alternating optimization of the precoding and the combining vectors takes place transparently at the CPU. Hence, for fixed combining vectors, the CPU solves the weighted sum MSE minimization problem
\begin{align} \label{eq:prob_centr}
\begin{array}{cl}
\displaystyle \min_{\{\w_{k}\}_{k \in \setK}} & \displaystyle \tr(\W^{\herm} \Phib \W) - 2 \Re \big[ \tr(\Omegab \H^{\herm} \W) \big] \\
\mathrm{s.t.} & \displaystyle \sum_{k \in \setK} \| \E_{b} \w_{k}\|^{2} \leq \rho_{\bs}, \quad \forall b \in \setB
\end{array}
\end{align}
where $\E_{b} \in \Real^{M \times B M}$ is a selection matrix such that $\E_{b} \w_{k} = \w_{b,k}$. For each UE~$k$, the first-order optimality condition of \eqref{eq:prob_centr} reads as
\begin{align} \label{eq:kkt_centr}
\nabla_{\w_{k}} \bigg( \sum_{\bar{k} \in \setK} \omega_{\bar{k}} \mse_{\bar{k}} + \sum_{b \in \setB} \lambda_{b} \bigg( \sum_{\bar{k} \in \setK} \|\E_{b} \w_{\bar{k}}\|^{2} - \rho_{\bs} \bigg) \bigg) = \0
\end{align}
where $\{\lambda_{b}\}_{b \in \setB}$ are the (coupled) dual variables related with the per-BS power constraints, which can be optimized, e.g., using the ellipsoid method. Finally, \eqref{eq:kkt_centr} yields the centralized precoding solution
\begin{align}
\label{eq:w_k} \w_{k} & = \omega_{k} \bigg( \Phib + \sum_{b \in \setB} \lambda_{b} \E_{b}^{\herm} \E_{b} \bigg)^{-1} \h_{k} \\
\label{eq:w_k_block} & = \omega_{k}
\begin{bmatrix}
\Phib_{1 1} + \lambda_{1} \I_{M}    & \hspace{-2mm} \ldots  & \hspace{-2mm} \Phib_{1 B} \\
\vdots                              & \hspace{-2mm} \ddots  & \hspace{-2mm} \vdots \\
\Phib_{1 B}^{\herm}                 & \hspace{-2mm} \ldots  & \hspace{-2mm} \Phib_{B B} + \lambda_{B} \I_{M}
\end{bmatrix}^{-1}
\begin{bmatrix}
\h_{1,k} \\
\vdots \\
\h_{B,k}
\end{bmatrix}.
\end{align}

The centralized precoding design is carried out as follows. First, each BS~$b$ acquires the channel matrices $\{\H_{b,k}\}_{k \in \setK}$ and forwards them to the CPU via backhaul signaling. Then, the CPU computes the aggregated precoding vectors $\{\w_{k}\}_{k \in \setK}$ as in \eqref{eq:w_k} together with the combining vectors $\{\v_{k}\}_{k \in \setK}$ as in \eqref{eq:v_k} by means of alternating optimization. Subsequently, it feeds back the BS-specific precoding vectors $\{\w_{b,k}\}_{k \in \setK}$ to each BS~$b$ via backhaul signaling. Lastly, each UE~$k$ acquires $\Psib_{k}$ in \eqref{eq:Psi_k} and the effective downlink channel $\g_{k}$, based on which it computes its combining vector $\v_{k}$ as in \eqref{eq:v_k}.

\subsection{Distributed Precoding Design via Backhaul Signaling} \label{sec:PROB_distr}

In the distributed precoding design, the BS-specific precoding vectors are computed locally by the BSs. Here, the alternating optimization of the precoding and the combining vectors takes place by means of iterative bi-directional training between the BSs and the UEs (see~\cite{Shi14,Kal18,Jay18,Tol19}). Hence, for fixed combining vectors, the BSs jointly solve the weighted sum MSE minimization problem
\begin{align} \label{eq:prob_distr}
\begin{array}{cl}
\displaystyle \min_{\{\w_{b,k}\}_{b \in \setB, k \in \setK}} & \displaystyle \tr(\W^{\herm} \Phib \W) - 2 \Re \big[ \tr(\Omegab \H^{\herm} \W) \big] \\
\mathrm{s.t.} & \displaystyle \sum_{k \in \setK} \|\w_{b,k}\|^{2} \leq \rho_{\bs}, \quad \forall b \in \setB.
\end{array}
\end{align}
For each BS~$b$ and for each UE~$k$, the first-order optimality condition of \eqref{eq:prob_distr} reads as
\begin{align} \label{eq:kkt_distr}
\nabla_{\w_{b,k}} \bigg( \sum_{\bar{k} \in \setK} \omega_{\bar{k}} \mse_{\bar{k}} + \sum_{\bar{b} \in \setB} \lambda_{\bar{b}} \bigg( \sum_{\bar{k} \in \setK} \|\w_{\bar{b},\bar{k}}\|^{2} - \rho_{\bs} \bigg) \bigg) = \0
\end{align}
where $\lambda_{b}$ has the same meaning as in \eqref{eq:kkt_centr} and can be optimized via bisection methods. Finally, \eqref{eq:kkt_distr} yields the distributed precoding solution\footnote{The equivalence between the centralized and the distributed precoding solutions in \eqref{eq:w_k} and \eqref{eq:w_bk}, respectively, is shown in Appendix~\ref{app:equiv} for the simple case of $B=2$~BSs, which can be extended to any value of $B$ by recursively applying the Schur complement.}
\begin{align} \label{eq:w_bk}
\w_{b,k} = (\Phib_{b b} + \lambda_{b} \I_{M})^{-1} (\omega_{k} \h_{b,k} - \xib_{b,k})
\end{align}
where we have defined
\begin{align} \label{eq:xi_bk}
\xib_{b,k} \triangleq \sum_{\bar{b} \in \setB \setminus \{b\}} \Phib_{b \bar{b}} \w_{\bar{b},k}.
\end{align}
Recall that the computation of $\{\w_{b,k}\}_{k \in \setK}$ by BS~$b$ requires the optimization of the dual variable $\lambda_{b}$ via bisection methods.\footnote{Observe that $\sum_{k \in \setK} \| \Phib_{b b}^{-1} (\omega_{k} \h_{b,k} - \xib_{b,k}) \| \leq \rho_{\bs}$ implies $\lambda_{b}=0$.} Building on the parallel optimization framework proposed in~\cite{Scu14}, the distributed precoding design can be implemented in an iterative best-response fashion~\cite{Kal18}. Focusing on UE~$k$, at each iteration~$i$, each BS~$b$ locally computes $\w_{b,k}$ as in \eqref{eq:w_bk} in parallel with the other BSs for a fixed $\xib_{b,k}$ (and, thus, for fixed $\{\w_{\bar{b},k}\}_{\bar{b} \in \setB \setminus \{b\}}$); then, each BS~$b$ updates its precoding vector as
\begin{align} \label{eq:w_bk_i}
\w_{b,k}^{(i)} = (1-\alpha) \w_{b,k}^{(i-1)} + \alpha \w_{b,k}
\end{align}
with $\alpha \in (0,1]$. In this context, the update in \eqref{eq:w_bk_i} is necessary to limit the variation of the precoding vectors between consecutive iterations, where the step size $\alpha$ must be chosen to strike the proper balance between convergence speed and accuracy. We refer to~\cite{Kal18,Scu14} for more details on the choice of $\alpha$ and on the convergence properties. 

\vspace{-2mm}
\begin{remark} \rm{
The vector $\xib_{b,k}$ in \eqref{eq:xi_bk} contains implicit information about the channel correlation between BS~$b$ and the other BSs and about the precoding vectors adopted by the latter for UE~$k$. The knowledge of such cross-term information at each BS~$b$ is required to iteratively adjust the distributed precoding solution so that it converges to its centralized counterpart described in Section~\ref{sec:PROB_centr}. In this regard, omitting $\xib_{b,k}$ from \eqref{eq:w_bk} yields the highly suboptimal \textit{local MMSE precoding}. Note that, while the effective uplink channels $\{\h_{b,k}\}_{k \in \setK}$ (which are also used to compute $\Phib_{b b}$) can be acquired locally by each BS~$b$ via uplink training, the acquisition of $\{\xib_{b,k}\}_{k \in \setK}$ calls for extensive CSI exchange among the BSs via backhaul signaling~\cite{Kal18}. In Section~\ref{sec:OTA}, we propose a practical scheme to implement the distributed precoding design that relies solely on OTA signaling.}
\end{remark} \vspace{-2mm}

\vspace{-2mm}

\vspace{-2mm}
\begin{remark} \rm{The computational complexity associated with the distributed precoding design after $i$ iterations is $\setO (i \delta B M^{3})$, with $\delta$ being the number of bisection steps per iteration; here, the term $M^3$ follows from the $(M \times M)$-dimensional matrix inversion in \eqref{eq:w_bk}. On the other hand, the computational complexity associated with the centralized precoding design described in Section~\ref{sec:PROB_centr} is $\setO (\delta B^{3} M^{3})$, where the term $B^{3} M^{3}$ follows from the $(B M \times B M)$-dimensional matrix inversion in \eqref{eq:w_k}. Hence, despite its iterative nature, the distributed precoding design brings a substantial computational complexity reduction as the total number of BS antennas in the network~$B M$ is usually very large in cell-free massive MIMO contexts.}
\end{remark} \vspace{-2mm}

The distributed precoding design via backhaul signaling is carried out as follows. First, for fixed combining vectors $\{\v_{k}\}_{k \in \setK}$, each BS~$b$ acquires the effective uplink channels $\{\h_{b,k}\}_{k \in \setK}$ and, by means of backhaul signaling, the vectors $\{\xib_{b,k}\}_{k \in \setK}$. Then, it computes its precoding vectors $\{\w_{b,k}\}_{k \in \setK}$ locally as in \eqref{eq:w_bk} and updates them as in \eqref{eq:w_bk_i}. Subsequently, each UE~$k$ acquires $\Psib_{k}$ in \eqref{eq:Psi_k} and the effective downlink channel $\g_{k}$, based on which it computes its combining vector $\v_{k}$ locally as in \eqref{eq:v_k}. This process is iterated until a predefined termination criterion is satisfied.

\section{Problem Formulation with Imperfect CSI} \label{sec:PROB_IMP}

In this section, we consider the centralized and the distributed precoding designs described in Sections~\ref{sec:PROB_centr}~and~\ref{sec:PROB_distr} under realistic pilot-aided CSI acquisition at both the BSs and the UEs (see Sections~\ref{sec:SM_ul}~and~\ref{sec:SM_dl}). Here, the precoding vectors $\{\w_{b,k}\}_{b \in \setB, k \in \setK}$ and the combining vectors $\{\v_{k}\}_{k \in \setK}$ are computed as the solutions of an estimated weighted sum MSE minimization problem with per-BS power constraints. For notational simplicity, and without loss of generality, we assume $\{\omega_{k} = 1\}_{k \in \setK}$.

\subsection{Centralized Precoding Design} \label{sec:PROB_IMP_centr}

In the centralized precoding design, the CPU computes the combining vectors and the aggregate precoding vectors for each UE~$k$ as
\begin{align}
\label{eq:v_k_centr_imp} \v_{k} & = \bigg( \sum_{\bar{k} \in \setK} \bigg( \sum_{b \in \setB} \hat{\H}_{b,k}^{\herm} \w_{b,\bar{k}} \bigg) \bigg( \sum_{b \in \setB} \w_{b,\bar{k}}^{\herm} \hat{\H}_{b,k} \bigg) + \sigma_{k}^{2} \I_{N} \bigg)^{-1} \sum_{b \in \setB} \hat{\H}_{b,k}^{\herm} \w_{b,k}, \\
\label{eq:w_k_centr_imp} \w_{k} & = \bigg( \sum_{\bar{k} \in \setK} \hat{\H}_{\bar{k}} \v_{\bar{k}} \v_{\bar{k}}^{\herm} \hat{\H}_{\bar{k}}^{\herm} + \sum_{b \in \setB} \lambda_{b} \E_{b}^{\herm} \E_{b} \bigg)^{-1} \hat{\H}_{k} \v_{k}
\end{align}
respectively, as part of the alternating optimization routine. Here, \eqref{eq:v_k_centr_imp} and \eqref{eq:w_k_centr_imp} are obtained from minimizing the sum MSE after replacing the channels $\{\H_{b,k}\}_{b \in \setB}$ with the estimated channels $\{\hat{\H}_{b,k}\}_{b \in \setB}$ (obtained as in \eqref{eq:H_bk_hat}) in \eqref{eq:MSE_k}, and are equal to \eqref{eq:v_k}~and~\eqref{eq:w_k}, respectively, for perfect channel estimation. The implementation of the centralized precoding design is formalized in Algorithm~\ref{alg:centr}, which is guaranteed to converge to a local optimum of the estimated sum MSE minimization problem (the same holds for Algorithm~\ref{alg:centr_it}). Note that such scheme is highly susceptible to imperfect channel estimation as it hinges on a single pilot-aided CSI acquisition (see Remark~\ref{rem:averaging}). Moreover, the amount of backhaul signaling for CSI exchange in step S.2 of Algorithm~\ref{alg:centr} scales with the number of BSs and UEs.

\begin{figure}[t!]
\begin{algorithm}[H]
\begin{algorithmic}
\begin{spacing}{1.25}
\STATE \hspace{-4mm} \textbf{Data:} Pilot matrices $\{\P_{k}\}_{k \in \setK}$ and pilot sequences $\{\p_{k}\}_{k \in \setK}$ ($\p_{k}$ can be the first column~of~$\P_{k}$).
\begin{itemize}[leftmargin=12mm]
\item[\texttt{(S.1)}] \textbf{UL:} Each UE~$k$ transmits the pilot matrix $\P_{k}$ (see $\X_{k}^{\ul}$ in \eqref{eq:X_k_ul}); each BS~$b$ receives~$\Y_{b}^{\ul}$~in~\eqref{eq:Y_b_ul}.
\item[\texttt{(S.2)}] Each BS~$b$ obtains $\{\hat{\H}_{b,k}\}_{k \in \setK}$ as in \eqref{eq:H_bk_hat} and forwards them to the CPU via backhaul signaling.
\item[\texttt{(S.3)}] The CPU computes the aggregated precoding vectors $\{\w_{k}\}_{k \in \setK}$ as in \eqref{eq:w_k_centr_imp} together with the combining vectors $\{\v_{k}\}_{k \in \setK}$ as in \eqref{eq:v_k_centr_imp} by means of alternating optimization.
\item[\texttt{(S.4)}] The CPU feeds back the BS-specific precoding vectors $\{\w_{b,k}\}_{k \in \setK}$ to each BS~$b$ via backhaul signaling.
\item[\texttt{(S.5)}] \textbf{DL:} Each BS~$b$ transmits a superposition of the pilot sequences $\{\p_{k}\}_{k \in \setK}$ after precoding them with the corresponding precoding vectors $\{\w_{b,k}\}_{k \in \setK}$ (see $\X_{b}^{\dl}$ in \eqref{eq:X_b_dl}); each UE~$k$ receives $\Y_{k}^{\dl}$ in \eqref{eq:Y_k_dl}.
\item[\texttt{(S.6)}] Each UE~$k$ computes its combining vector $\v_{k}$ as in \eqref{eq:v_k_distr_imp}.
\end{itemize}
\vspace{-5mm}
\end{spacing}
\end{algorithmic}
\caption{(Centralized)} \label{alg:centr}
\end{algorithm}
\vspace{-12mm}
\end{figure}

\subsection{Distributed Precoding Design via Backhaul Signaling} \label{sec:PROB_IMP_distr}

In the distributed precoding design, after the downlink pilot-aided channel estimation phase, each UE~$k$ obtains
\begin{align}
\label{eq:Psi_k_est} \frac{1}{\tau} \Y_{k}^{\dl} (\Y_{k}^{\dl})^{\herm} = \ & \Psib_{k} + \frac{1}{\tau} \sum_{\substack{\bar{k},j \in \setK \\ \bar{k} \neq j}} \bigg( \sum_{b \in \setB} \H_{b,k}^{\herm} \w_{b,\bar{k}} \bigg) \bigg( \sum_{b \in \setB} \w_{b,j}^{\herm} \H_{b,k} \bigg) (\p_{\bar{k}}^{\herm} \p_{j}) + \N_{k}^{\dl}
\end{align}
with $\Y_{k}^{\dl}$ and $\Psib_{k}$ defined in \eqref{eq:Y_k_dl} and \eqref{eq:Psi_k}, respectively, and
\begin{align}
\label{eq:N_k} \N_{k}^{\dl} \triangleq \frac{1}{\tau} \bigg( \sum_{b \in \setB} & \sum_{\bar{k} \in \setK} \bigg( \H_{b,k}^{\herm} \w_{b,\bar{k}} \p_{\bar{k}}^{\herm} (\Z_{k}^{\dl})^{\herm} + \Z_{k}^{\dl} \p_{\bar{k}} \w_{b,\bar{k}}^{\herm} \H_{b,k} \bigg) + \Z_{k}^{\dl} (\Z_{k}^{\dl})^{\herm} \bigg).
\end{align}
Here, perfect channel estimation would imply that:
\begin{itemize}[leftmargin=6mm]
\item[\textit{i)}] The pilot contamination in the second term of \eqref{eq:Psi_k_est} is eliminated;
\item[\textit{ii)}] As $\tau \to \infty$, we have that $\N_{k}^{\dl} \to \sigma_{k}^{2} \I_{N}$.
\end{itemize}
Hence, UE~$k$ can use \eqref{eq:Psi_k_est} as an estimate of $\Psib_{k} + \sigma_{k}^{2} \I_{N}$ and, consequently, it can obtain an estimate of $\mse_{k}$ in \eqref{eq:MSE_k_v} as
\begin{align} \label{eq:MSE_k_v_imp}
\mse_{k} \simeq \frac{1}{\tau} \v_{k}^{\herm} \Y_{k}^{\dl} (\Y_{k}^{\dl})^{\herm} \v_{k} - \frac{2}{\tau} \Re [\v_{k}^{\herm} \Y_{k}^{\dl} \p_{k}] + 1.
\end{align}
Finally, each UE~$k$ can compute its combining vector $\v_{k}$ locally as
\begin{align} \label{eq:v_k_distr_imp}
\v_{k} & = \big( \Y_{k}^{\dl} (\Y_{k}^{\dl})^{\herm} \big)^{-1} \Y_{k}^{\dl} \p_{k}
\end{align}
which is equal to \eqref{eq:v_k} for perfect channel estimation.

On the other hand, for the computation of the precoding vectors, let us define $\Y^{\ulA} \triangleq [(\Y_{1}^{\ulA})^{\tran}, \ldots, (\Y_{K}^{\ulA})^{\tran}]^{\tran} \in \Compl^{B M \times \tau}$ and $\P \triangleq [\p_{1}, \ldots, \p_{K}] \in \Compl^{\tau \times K}$. The following steps describe how the cross-term information can be expressed in terms of the receive signals at the BSs in the uplink pilot-aided channel estimation phase. For each BS pair $b$ and $\bar{b}$, we have the following relation:
\begin{align} \label{eq:Phi_bb_est}
\frac{1}{\tau \beta^{\ulA}} \Y_{b}^{\ulA} (\Y_{\bar{b}}^{\ulA})^{\herm} = \Phib_{b \bar{b}} + \frac{1}{\tau} \sum_{\substack{k,\bar{k} \in \setK \\ k \neq \bar{k}}} \h_{b,k} \h_{\bar{b},\bar{k}}^{\herm} (\p_{k}^{\herm} \p_{\bar{k}}) + \N_{b \bar{b}}^{\ulA}
\end{align}
with $\Y_{b}^{\ulA}$ defined in \eqref{eq:Y_b_ul1} and
\begin{align} \label{eq:N_bb}
\N_{b \bar{b}}^{\ulA} \triangleq \frac{1}{\tau} \bigg( \frac{1}{\sqrt{\beta^{\ulA}}} \sum_{k \in \setK} \bigg( \h_{b,k} \p_{k}^{\herm} (\Z_{\bar{b}}^{\ulA})^{\herm} & + \Z_{b}^{\ulA} \p_{k} \h_{\bar{b},k}^{\herm} \bigg) + \frac{1}{\beta^{\ulA}} \Z_{b}^{\ulA} (\Z_{\bar{b}}^{\ulA})^{\herm} \bigg).
\end{align}
Note that $\Y_{b}^{\ulA}$ is not available at BS~$\bar{b} \neq b$. Here, perfect channel estimation would imply that:
\begin{itemize}[leftmargin=6mm]
\item[\textit{i)}] The pilot contamination in the second term of \eqref{eq:Phi_bb_est} is eliminated;
\item[\textit{ii)}] As $\tau \to \infty$, we have that $\N_{b \bar{b}}^{\ulA} \to \0$ if $\bar{b} \neq b$ and $\N_{b b}^{\ulA} \to \frac{\sigma_{b}^{2}}{\beta^{\ulA}} \I_{M}$.
\end{itemize}
Hence, \eqref{eq:Phi_bb_est} can be intended as an estimate of $\Phib_{b \bar{b}}$ if $\bar{b} \neq b$ or of $\Phib_{b b} + \frac{\sigma_{b}^{2}}{\beta^{\ulA}} \I_{M}$ if $\bar{b} = b$ and, consequently, $\frac{1}{\tau \beta^{\ulA}} \Y^{\ulA} (\Y^{\ulA})^{\herm}$ can be intended as an estimate of $\Phib + \frac{1}{\beta^{\ulA}} \sum_{b \in \setB} \sigma_{b}^{2} \E_{b}^{\herm} \E_{b}$. This can be exploited to write the estimated sum MSE as
\begin{align}
\nonumber \sum_{k \in \setK} \mse_{k} \simeq \frac{1}{\tau \beta^{\ulA}} \tr \bigg( \W^{\herm} \bigg( & \Y^{\ulA} (\Y^{\ulA} )^{\herm} - \tau \sum_{b \in \setB} \sigma_{b}^{2} \E_{b}^{\herm} \E_{b} \bigg) \W \bigg) \\
\label{eq:sum_MSE_imp} & - \frac{2}{\tau \sqrt{\beta^{\ulA}}} \Re \big[ \tr \big( \P^{\herm} (\Y^{\ulA})^{\herm} \W \big) \big] + \sum_{k \in \setK} \sigma_{k}^{2} \| \v_{k} \|^{2} + K
\end{align}
where the term $- \frac{1}{\beta^{\ulA}} \sum_{b \in \setB} \sigma_{b}^{2} \E_{b}^{\herm} \E_{b}$ removes the noise bias from the estimation of $\Phib$. For fixed combining vectors, the BSs jointly solve the estimated sum MSE minimization problem
\begin{align} \label{eq:prob_distr_imp}
\hspace{-1mm} \begin{array}{cl}
\displaystyle \min_{\{\w_{b,k}\}_{b \in \setB, k \in \setK}} & \displaystyle \tr \bigg( \W^{\herm} \bigg( \Y^{\ulA} (\Y^{\ulA})^{\herm} - \tau \sum_{b \in \setB} \sigma_{b}^{2} \E_{b}^{\herm} \E_{b} \bigg) \W \bigg) - 2 \sqrt{\beta^{\ulA}} \Re \big[ \tr \big( \P^{\herm} (\Y^{\ulA})^{\herm} \W \big) \big] \\
\mathrm{s.t.} & \displaystyle \sum_{k \in \setK} \|\w_{b,k}\|^{2} \leq \rho_{\bs}, \quad \forall b \in \setB.
\end{array}
\end{align}
\begin{figure}[t!]
\begin{algorithm}[H]
\begin{algorithmic}
\begin{spacing}{1.25}
\STATE \hspace{-4mm} \textbf{Data:} Pilot sequences $\{\p_{k}\}_{k \in \setK}$.
\STATE \hspace{-4mm} \textbf{Initialization:} Each BS~$b$ initializes its precoding vectors $\{\w_{b,k}^{(0)}\}_{k \in \setK}$; set $i=0$.
\STATE \hspace{-4mm} \textbf{Until} a predefined termination criterion is satisfied, \textbf{do:}
\begin{itemize}[leftmargin=12mm]
\item[\texttt{(S.0)}] $i \leftarrow i+1$.
\item[\texttt{(S.1)}] \textbf{DL:} Each BS~$b$ transmits a superposition of the pilot sequences $\{\p_{k}\}_{k \in \setK}$ after precoding them with the corresponding precoding vectors $\{\w_{b,k}\}_{k \in \setK}$ (see $\X_{b}^{\dl}$ in \eqref{eq:X_b_dl}); each UE~$k$ receives $\Y_{k}^{\dl}$ in \eqref{eq:Y_k_dl}.
\item[\texttt{(S.2)}] Each UE~$k$ computes its combining vector $\v_{k}$ as in \eqref{eq:v_k_distr_imp}.
\item[\texttt{(S.3)}] \textbf{UL-1:} Each UE~$k$ transmits its pilot sequence $\p_{k}$ after precoding it with its combining vector $\v_{k}$ (see $\X_{k}^{\ulA}$ in \eqref{eq:X_k_ul1}); each BS~$b$ receives $\Y_{b}^{\ulA}$ in \eqref{eq:Y_b_ul1}.
\item[\texttt{(S.4)}] For each UE~$k$, each BS~$b$ acquires $\big\{ (\Y_{\bar{b}}^{\ulA})^{\herm} \w_{\bar{b},k} \big\}_{\bar{b} \in \setB \setminus \{b\}}$ from the other BSs via backhaul signaling.
\item[\texttt{(S.5)}] For each UE~$k$, each BS~$b$ computes its precoding vectors $\{\w_{b,k}\}_{k \in \setK}$ as in \eqref{eq:w_bk_imp} and updates them as in \eqref{eq:w_bk_i}.
\end{itemize}
\STATE \hspace{-4mm} \textbf{End}
\vspace{-5mm}
\end{spacing}
\end{algorithmic}
\caption{(Distributed--backhaul)} \label{alg:distr_bh}
\end{algorithm}
\vspace{-12mm}
\end{figure}
Finally, for each BS~$b$ and for each UE~$k$, the first-order optimality condition of \eqref{eq:prob_distr_imp} yields the distributed precoding solution
\begin{align} \label{eq:w_bk_imp}
\w_{b,k} = \big( \Y_{b}^{\ulA} & (\Y_{b}^{\ulA})^{\herm} + \tau (\beta^{\ulA} \lambda_{b} - \sigma_{b}^{2}) \I_{M} \big)^{-1} \Y_{b}^{\ulA} \bigg( \sqrt{\beta^{\ulA}} \p_{k} - \sum_{\bar{b} \in \setB \setminus \{b\}} (\Y_{\bar{b}}^{\ulA})^{\herm} \w_{\bar{b},k} \bigg)
\end{align}
which is equal to \eqref{eq:w_bk} for perfect channel estimation, and where the term $- \tau \sigma_{b}^{2} \I_{M}$ in the inverse matrix removes the noise bias from the estimation of $\tau \beta^{\ulA} \Phib_{bb}$ (the same holds for \eqref{eq:w_k_imp_it_block}, \eqref{eq:xi_est}, and \eqref{eq:w_bk_imp_ota}). To compute $\w_{b,k}$ as in \eqref{eq:w_bk_imp}, BS~$b$ needs to acquire the term $(\Y_{\bar{b}}^{\ulA})^{\herm} \w_{\bar{b},k} \in \Compl^{\tau \times 1}$ from each BS~$\bar{b} \in \setB \setminus \{b\}$ via backhaul signaling, as described in~\cite{Kal18}. The iterative implementation of the distributed precoding design via backhaul signaling is formalized in Algorithm~\ref{alg:distr_bh}, whose convergence to a local optimum of the estimated sum MSE minimization problem is guaranteed by the proper choice of the step size $\alpha$ in \eqref{eq:w_bk_i} (the same holds for Algorithm~\ref{alg:distr_ota}). Here, suitable termination criteria can be, for instance, $i=i^{\max}$, where $i^{\max}$ is the maximum number of iterations (fixed to comply with some latency constraints or adapted to the duration of the scheduling block), $|R^{(i)}-R^{(i-1)}| \leq \epsilon$, or $\|\W^{(i)} - \W^{(i-1)}\|_{\rmF}^{2} \leq \epsilon$. These observations also apply to Algorithms~\ref{alg:centr_it}~and~\ref{alg:distr_ota} in the following.

\vspace{-2mm}
\begin{remark} \rm{
The amount of backhaul signaling for CSI exchange in step S.4 of Algorithm~\ref{alg:distr_bh} scales not only with the pilot length $\tau$ and the number of bi-directional training iterations, but also with the number of BSs~$B$ and the number of UEs~$K$ since the cross terms are specific for each BS-UE pair. This becomes burdensome in cell-free massive MIMO contexts due to the large number of BSs and UEs involved in the joint processing. In addition, the CSI exchange among the BSs via backhaul signaling does not occur instantaneously.\footnote{Without loss of generality, one can express the delay introduced by the backhaul into the CSI exchange in terms of number of bi-directional training iterations. In our numerical results in Section~\ref{sec:NUM}, we assume that such delay amounts to one bi-directional training iteration.} Therefore, each BS must rely on outdated CSI from the other BSs, which can significantly degrade the performance of the distributed precoding design (as demonstrated in~\cite{Kal18}). In Section~\ref{sec:OTA}, we propose a practical scheme that allows each BS to acquire the missing cross-term information via OTA signaling, which entirely eliminates the need for backhaul signaling for CSI exchange among the BSs.}
\end{remark} \vspace{-2mm}

\vspace{-2mm}

\vspace{-2mm}
\begin{remark} \label{rem:averaging} \rm{
As detailed in~\cite{Kal18}, using \eqref{eq:Phi_bb_est} as a surrogate of $\Phib_{b b}$ provides improved robustness against pilot contamination with respect to estimating each UE channel explicitly. Consequently, the distributed precoding design is less sensitive to pilot contamination than the centralized precoding design described in Section~\ref{sec:PROB_IMP_centr}. Even in absence of pilot contamination, due to its iterative nature that involves several pilot-aided CSI acquisitions, the distributed precoding design is more robust to noisy channel estimation than its centralized counterpart (which hinges on a single pilot-aided CSI acquisition). In this regard, it is straightforward to observe that the precoding vector update at iteration~$i$, i.e., $\w_{b,k}^{(i)}$ defined in \eqref{eq:w_bk_i}, can be expressed as a weighted average of $i$ precoding vectors computed as in \eqref{eq:w_bk_imp} based on as many channel estimations with independent AWGN realizations. Hence, the update in \eqref{eq:w_bk_i} produces a beneficial averaging of the channel estimation noise that reflects positively on the sum-rate performance. The robustness of the distributed precoding design against both pilot contamination and channel estimation noise is highlighted in our numerical results in Section~\ref{sec:NUM}.}
\end{remark} \vspace{-2mm}

For comparative purposes, in the next section, we present a centralized precoding design with iterative bi-directional training between the BSs (which communicate with the CPU via backhaul signaling) and the UEs. Similarly to the distributed precoding design, this scheme involves pilot-aided CSI acquisitions at each bi-directional training iteration and thus overcomes the main drawback of the centralized precoding design described in Section~\ref{sec:PROB_IMP_centr}.

\subsection{Centralized Precoding Design with Iterative Bi-Directional Training} \label{sec:PROB_IMP_centr_it}

In the centralized precoding design with iterative bi-directional training, the aggregated precoding vectors are computed by the CPU and the BS-specific precoding vectors are fed back to the corresponding BSs via backhaul signaling. Unlike the centralized precoding design in Algorithm~\ref{alg:centr}, which hinges on a single pilot-aided CSI acquisition, the alternating optimization of the precoding and the combining vectors takes place by means of iterative bi-directional training between the CPU and the UEs through the BSs (as in the distributed precoding design in Algorithm~\ref{alg:distr_bh}). Hence, for fixed combining vectors, the CPU solves the estimated sum MSE minimization problem
\begin{align} \label{eq:prob_centr_imp}
\hspace{-3mm} \begin{array}{cl}
\displaystyle \min_{\{\w_{k}\}_{k \in \setK}} & \displaystyle \tr \bigg( \W^{\herm} \bigg( \Y^{\ulA} (\Y^{\ulA})^{\herm} - \tau \sum_{b \in \setB} \sigma_{b}^{2} \E_{b}^{\herm} \E_{b} \bigg) \W \bigg) \! - \! 2 \sqrt{\beta^{\ulA}} \Re \big[ \tr \big( \P^{\herm} (\Y^{\ulA})^{\herm} \W \big) \big] \\
\mathrm{s.t.} & \displaystyle \sum_{k \in \setK} \| \E_{b} \w_{k}\|^{2} \leq \rho_{\bs}, \quad \forall b \in \setB.
\end{array}
\end{align}
For each UE~$k$, the first-order optimality condition of \eqref{eq:prob_centr_imp} yields the centralized precoding solution\footnote{The equivalence between the centralized and the distributed precoding solutions in \eqref{eq:w_k_imp_it} and \eqref{eq:w_bk_imp}, respectively, can be shown in the same way as in the case with perfect CSI (see Section~\ref{sec:PROB}).}
\begin{figure}[t!]
\begin{algorithm}[H]
\begin{algorithmic}
\begin{spacing}{1.25}
\STATE \hspace{-4mm} \textbf{Data:} Pilot sequences $\{\p_{k}\}_{k \in \setK}$.
\STATE \hspace{-4mm} \textbf{Initialization:} The CPU initializes the aggregated precoding vectors $\{\w_{k}^{(0)}\}_{k \in \setK}$; set $i=0$.
\STATE \hspace{-4mm} \textbf{Until} a predefined termination criterion is satisfied, \textbf{do:}
\begin{itemize}[leftmargin=12mm]
\item[\texttt{(S.0)}] $i \leftarrow i+1$.
\item[\texttt{(S.1)}] The CPU feeds back the BS-specific precoding vectors $\{\w_{b,k}\}_{k \in \setK}$ to each BS~$b$ via backhaul signaling.
\item[\texttt{(S.2)}] \textbf{DL:} Each BS~$b$ transmits a superposition of the pilot sequences $\{\p_{k}\}_{k \in \setK}$ after precoding them with the corresponding precoding vectors $\{\w_{b,k}\}_{k \in \setK}$ (see $\X_{b}^{\dl}$ in \eqref{eq:X_b_dl}); each UE~$k$ receives $\Y_{k}^{\dl}$ in \eqref{eq:Y_k_dl}.
\item[\texttt{(S.3)}] Each UE~$k$ computes its combining vector $\v_{k}$ as in \eqref{eq:v_k_distr_imp}.
\item[\texttt{(S.4)}] \textbf{UL-1:} Each UE~$k$ transmits its pilot sequence $\p_{k}$ after precoding it with its combining vector $\v_{k}$ (see $\X_{k}^{\ulA}$ in \eqref{eq:X_k_ul1}); each BS~$b$ receives $\Y_{b}^{\ulA}$ in \eqref{eq:Y_b_ul1}.
\item[\texttt{(S.5)}] Each BS~$b$ forwards $\Y_{b}^{\ulA}$ to the CPU via backhaul signaling.
\item[\texttt{(S.6)}] The CPU computes the precoding vectors $\{\w_{k}\}_{k \in \setK}$ as in \eqref{eq:w_k_imp_it}.
\end{itemize}
\STATE \hspace{-4mm} \textbf{End}
\vspace{-5mm}
\end{spacing}
\end{algorithmic}
\caption{(Centralized--iterative)} \label{alg:centr_it}
\end{algorithm}
\vspace{-12mm}
\end{figure}
\begin{align}
\label{eq:w_k_imp_it} \w_{k} \! & = \! \sqrt{\beta^{\ulA}} \bigg( \Y^{\ulA} (\Y^{\ulA})^{\herm} + \tau \sum_{b \in \setB} (\beta^{\ulA} \lambda_{b} - \sigma_{b}^{2}) \E_{b}^{\herm} \E_{b} \bigg)^{-1} \Y^{\ulA} \p_{k} \\
\label{eq:w_k_imp_it_block} & = \! \sqrt{\beta^{\ulA}} \!
\begin{bmatrix}
\! \Y_{1}^{\ulA} (\Y_{1}^{\ulA})^{\herm} \! + \! \tau (\beta^{\ulA} \lambda_{1} \! - \! \sigma_{1}^{2}) \I_{M} & \hspace{-3mm} \ldots \hspace{-3mm} & \Y_{1}^{\ulA} (\Y_{B}^{\ulA})^{\herm} \\
\vdots & \hspace{-3mm} \ddots \hspace{-3mm} & \vdots \\
\Y_{B}^{\ulA} (\Y_{1}^{\ulA})^{\herm} & \hspace{-3mm} \ldots \hspace{-3mm} & \Y_{B}^{\ulA} (\Y_{B}^{\ulA})^{\herm} \! + \! \tau (\beta^{\ulA} \lambda_{B} \! - \! \sigma_{B}^{2}) \I_{M} \!
\end{bmatrix}^{-1} \! \!
\begin{bmatrix}
\! \Y_{1}^{\ulA} \! \\
\vdots \\
\! \Y_{B}^{\ulA} \!
\end{bmatrix} \! \p_{k}
\end{align}
which is equal to \eqref{eq:w_k} for perfect channel estimation. The implementation of the centralized precoding design with iterative bi-directional training is formalized in Algorithm~\ref{alg:centr_it}. This scheme is used for comparative purposes in our numerical results in Section~\ref{sec:NUM}; however, the high computational complexity resulting from the centralized precoding design combined with the cumbersome backhaul signaling between the BSs and the CPU make its implementation highly impractical.

\section{Distributed Precoding Design via OTA Signaling} \label{sec:OTA}

In this section, we propose a novel OTA signaling scheme that entirely eliminates the need for backhaul signaling for CSI exchange among the BSs and, hence, overcomes the practical limitations of the distributed precoding design described in Section~\ref{sec:PROB_IMP_distr}. To this end, we introduce a \textit{new uplink signaling resource} together with a \textit{new CSI combining mechanism} that complement the existing uplink and downlink signaling described in Sections~\ref{sec:SM_ul}~and~\ref{sec:SM_dl}, respectively. This allows each BS to acquire the missing cross-term information necessary for the distributed precoding design over the air rather than via extensive backhaul signaling among the BSs.

After the uplink and the downlink pilot-aided channel estimation phases, each BS~$b$ obtains an estimate of $\xib_{b,k}$ in \eqref{eq:xi_bk} as described next. Upon computing its combining vector, each UE~$k$ synchronously retransmits $\Y_{k}^{\dl}$ in \eqref{eq:Y_k_dl} after multiplying it by the rank-1 matrix $\v_{k} \v_{k}^{\herm}$, i.e.,
\begin{align} \label{eq:X_k_ul2}
\X_{k}^{\ulB} \triangleq \sqrt{\beta^{\ulB}} \v_{k} \v_{k}^{\herm} \Y_{k}^{\dl} \in \Compl^{N \times \tau}
\end{align}
where the power scaling factor $\beta^{\ulB}$ (equal for all the UEs) ensures that $\X_{k}^{\ulB}$ complies with the UE transmit power constraint (see Section~\ref{sec:OTA_power} for more details on the choice of $\beta^{\ulB}$). More specifically, each UE~$k$ uses its combining vector $\v_{k}$ to combine $\Y_{k}^{\dl}$ and then transmits the combined signal $\v_{k}^{\herm} \Y_{k}^{\dl}$ using again $\v_{k}$ as precoder, which does not increase the computational complexity at the UE. Then, the receive signal at BS~$b$ is given by
\begin{align}
\Y_{b}^{\ulB} & \triangleq \sum_{k \in \setK} \H_{b,k} \X_{k}^{\ulB} + \Z_{b}^{\ulB} \\
\label{eq:Y_b_ul2} & = \sqrt{\beta^{\ulB}} \sum_{k \in \setK} \h_{b,k} \v_{k}^{\herm} \bigg( \sum_{\bar{b} \in \setB} \sum_{\bar{k} \in \setK} \H_{\bar{b},k}^{\herm} \w_{\bar{b},\bar{k}} \p_{\bar{k}}^{\herm} + \Z_{k}^{\dl} \bigg) + \Z_{b}^{\ulB} \in \Compl^{M \times \tau}
\end{align}
where $\Z_{b}^{\ulB} \in \Compl^{M \times \tau}$ is the AWGN term at BS~$b$ with elements distributed as $\setC \setN (0, \sigma_{b}^{2})$. At this stage, it is easy to observe that $\Y_{b}^{\ulB}$ in \eqref{eq:Y_b_ul2} contains useful information about the channel correlation between BS~$b$ and the other BSs and about the precoding vectors adopted by the latter (which is necessary for the local computation of the precoding vectors). By means of this new uplink signaling resource, each BS~$b$ obtains
\begin{align} \label{eq:xi_est_1}
\frac{1}{\tau \sqrt{\beta^{\ulB}}} \Y_{b}^{\ulB} \p_{k} = \sum_{\bar{b} \in \setB} \Phib_{b \bar{b}} \w_{\bar{b},k} + \frac{1}{\tau} \sum_{\bar{k} \in \setK \setminus \{k\}} \sum_{\bar{b} \in \setB} \Phib_{b \bar{b}} \w_{\bar{b},\bar{k}} \p_{\bar{k}}^{\herm} \p_{k} + \n_{b,k}^{\ulB}
\end{align}
where we have defined \vspace{-1mm}
\begin{align} \label{eq:n_bk_ul2}
\n_{b,k}^{\ulB} \triangleq \frac{1}{\tau} \bigg( \sum_{\bar{k} \in \setK} \h_{b,\bar{k}} \v_{\bar{k}}^{\herm} \Z_{\bar{k}}^{\dl} + \frac{1}{\sqrt{\beta^{\ulB}}} \Z_{b}^{\ulB} \bigg) \p_{k}.
\end{align}
\noindent Here, perfect channel estimation would imply that:
\begin{itemize}[leftmargin=6mm]
\item[\textit{i)}] The pilot contamination in the second term of \eqref{eq:xi_est_1} is eliminated;
\item[\textit{ii)}] As $\tau \to \infty$, the noise term $\n_{b,k}^{\ulB}$ in \eqref{eq:n_bk_ul2} is eliminated.
\end{itemize}
Therefore, BS~$b$ can use \eqref{eq:xi_est_1} as an estimate of $\sum_{\bar{b} \in \setB} \Phib_{b \bar{b}} \w_{\bar{b},k}$. Then, each BS~$b$ can obtain an estimate of $\xib_{b,k}$ in \eqref{eq:xi_bk} by suitably combining the uplink signaling resources $\Y_{b}^{\ulA}$ and $\Y_{b}^{\ulB}$ as
\begin{align}
\nonumber & \frac{1}{\tau} \bigg( \frac{1}{\sqrt{\beta^{\ulB}}} \Y_{b}^{\ulB} \p_{k} - \frac{1}{\beta^{\ulA}} \big( \Y_{b}^{\ulA} (\Y_{b}^{\ulA})^{\herm} - \tau \sigma_{b}^{2} \I_{M} \big) \w_{b,k} \bigg) \\
\nonumber & \hspace{1mm} = \xib_{b,k} \! + \! \frac{1}{\tau} \bigg( \sum_{\bar{k} \in \setK \setminus \{k\}} \sum_{\bar{b} \in \setB} \Phib_{b \bar{b}} \w_{\bar{b},\bar{k}} \p_{\bar{k}}^{\herm} \p_{k} \! - \! \! \sum_{\substack{\bar{k},j \in \setK \\ \bar{k} \neq j}} \h_{b,\bar{k}} \h_{b,j}^{\herm} \w_{b,k} \p_{\bar{k}}^{\herm} \p_{j} \bigg) \! + \! \n_{b,k}^{\ulB} \! + \! \bigg( \frac{\sigma_{b}^{2}}{\beta^{\ulA}} \I_{M} \! - \! \N_{b b}^{\ulA} \bigg) \w_{b,k}
\end{align}

\vspace{-8mm}

\begin{align}
\label{eq:xi_est}
\end{align}

\vspace{-1mm}

\noindent (recall that, as $\tau \to \infty$, we have that $\N_{b b}^{\ulA} \to \frac{\sigma_{b}^{2}}{\beta^{\ulA}} \I_{M}$). In practice, the missing cross-term information is obtained by removing the local estimate of $\Phib_{bb} \w_{b,k}$, where the precoding vector is from the previous iteration, from \eqref{eq:xi_est_1}. Finally, for each BS~$b$ and for each UE~$k$, the distributed precoding solution via OTA signaling is obtained as \vspace{-2mm}
\begin{align}
\nonumber \w_{b,k} = \big( \Y_{b}^{\ulA} (\Y_{b}^{\ulA})^{\herm} & + \tau (\beta^{\ulA} \lambda_{b} - \sigma_{b}^{2}) \I_{M} \big)^{-1} \\
\label{eq:w_bk_imp_ota} & \times \bigg( \Y_{b}^{\ulA} \big( \sqrt{\beta^{\ulA}} \p_{k} + (\Y_{b}^{\ulA})^{\herm} \w_{b,k} \big) - \frac{\beta^{\ulA}}{\sqrt{\beta^{\ulB}}}\Y_{b}^{\ulB} \p_{k} - \tau \sigma_{b}^{2} \w_{b,k} \bigg)
\end{align}
which is equal to \eqref{eq:w_bk} for perfect channel estimation. The iterative implementation of the distributed precoding design via OTA signaling is formalized in Algorithm~\ref{alg:distr_ota} (see also Figure~\ref{fig:ota}). It is worth noting that the downlink pilot-aided channel estimation phase assumes a new importance in the context of this scheme: in fact, in addition to enabling the optimization of the combining vectors at the UEs, it has a crucial role in allowing the BSs to obtain the missing cross-term information over the air. In our recent work~\cite{Atz20}, the proposed OTA signaling mechanism has been adapted for the uplink scenario to enable distributed joint receiver design. Lastly, in presence of hybrid analog-digital precoding, the proposed distributed precoding design via OTA signaling can be used to jointly optimize the digital beamformers, whereas the analog beamformers would need to be computed locally at each BS.

\begin{figure}[t!]
\begin{algorithm}[H]
\begin{algorithmic}
\begin{spacing}{1.25}
\STATE \hspace{-4mm} \textbf{Data:} Pilot sequences $\{\p_{k}\}_{k \in \setK}$.
\STATE \hspace{-4mm} \textbf{Initialization:} Each BS~$b$ initializes its precoding vectors $\{\w_{b,k}^{(0)}\}_{k \in \setK}$; set $i=0$.
\STATE \hspace{-4mm} \textbf{Until} a predefined termination criterion is satisfied, \textbf{do:}
\begin{itemize}[leftmargin=12mm]
\item[\texttt{(S.0)}] $i \leftarrow i+1$.
\item[\texttt{(S.1)}] \textbf{DL:} Each BS~$b$ transmits a superposition of the pilot sequences $\{\p_{k}\}_{k \in \setK}$ after precoding them with the corresponding precoding vectors $\{\w_{b,k}\}_{k \in \setK}$ (see $\X_{b}^{\dl}$ in \eqref{eq:X_b_dl}); each UE~$k$ receives $\Y_{k}^{\dl}$ in \eqref{eq:Y_k_dl}.
\item[\texttt{(S.2)}] Each UE~$k$ computes its combining vector $\v_{k}$ as in \eqref{eq:v_k_distr_imp}.
\item[\texttt{(S.3)}] \textbf{UL-1:} Each UE~$k$ transmits its pilot sequence $\p_{k}$ after precoding it with its combining vector $\v_{k}$ (see $\X_{k}^{\ulA}$ in \eqref{eq:X_k_ul1}); each BS~$b$ receives $\Y_{b}^{\ulA}$ in \eqref{eq:Y_b_ul1}.
\item[\texttt{(S.4)}] \textbf{UL-2:} Each UE~$k$ transmits $\Y_{k}^{\dl}$ after precoding it with the rank-1 matrix $\v_{k} \v_{k}^{\herm}$ (see $\X_{k}^{\ulB}$ in \eqref{eq:X_k_ul2}); each BS~$b$ receives $\Y_{b}^{\ulB}$ in \eqref{eq:Y_b_ul2}.
\item[\texttt{(S.5)}] For each UE~$k$, each BS~$b$ computes its precoding vectors $\{\w_{b,k}\}_{k \in \setK}$ as in \eqref{eq:w_bk_imp} and updates them as in \eqref{eq:w_bk_i}.
\end{itemize}
\STATE \hspace{-4mm} \textbf{End}
\vspace{-5mm}
\end{spacing}
\end{algorithmic}
\caption{(Distributed--OTA)} \label{alg:distr_ota}
\end{algorithm}
\vspace{-12mm}
\end{figure}

\vspace{-2mm}
\begin{remark} \rm{In the distributed precoding design via OTA signaling, the CSI exchange among the BSs via backhaul signaling is entirely replaced by the new uplink signaling resource UL-2 (see step S.4 of Algorithm~\ref{alg:distr_ota}), with clear advantages in terms of scalability and flexibility. Remarkably, the proposed OTA signaling mechanism allows each BS~$b$ to recover the cross-term information for all the UEs from the same receive signal $\Y_{b}^{\ulB}$ (i.e., by correlating the latter with} the UE-specific pilot sequence as in \eqref{eq:xi_est_1}) rather than by exchanging cross terms specific for each BS-UE pair. Hence, the amount of OTA signaling does not scale with the number of BSs~$B$ or the number of UEs~$K$ (unlike the backhaul signaling associated with Algorithm~\ref{alg:distr_bh}), and depends only on the pilot length~$\tau$ and on the number of bi-directional training iterations. A further advantage of eliminating the CSI exchange via backhaul signaling is that more backhaul resources can be dedicated to the UE-specific data sharing, which, in turn, enables more BSs to cooperate in the joint transmission~\cite{Lee12}. This is crucial for the practical implementation of cell-free massive MIMO, where the number of cooperating BSs can be very large and even extend to the whole network. In addition, the delays introduced by the backhaul in the exchange of the cross-term information are eliminated. These practical benefits come at the cost of extra uplink signaling overhead per bi-directional training iteration with respect to Algorithm~\ref{alg:distr_bh} (where only DL and UL-1 are present). However, we note that the impact of the extra signaling overhead (and the corresponding performance loss) becomes negligible for sufficiently large scheduling blocks~\cite{Tol19}, as detailed in Section~\ref{sec:OTA_frame} and as shown in our numerical results in Section~\ref{sec:NUM}.
\end{remark} \vspace{-2mm}

In the following, we discuss two relevant implementation aspects of the proposed distributed precoding design via OTA signaling, namely: \textit{i)} how the OTA signaling can be integrated into the 5G 3GPP NR frame/slot structure (in Section~\ref{sec:OTA_frame}); and \textit{ii)} how the uplink training can be implemented in compliance with the transmit power constraint of the UEs (in Section~\ref{sec:OTA_power}).

\subsection{Implementing the OTA Signaling in 5G 3GPP NR} \label{sec:OTA_frame}

The distributed precoding design schemes described in this paper heavily rely on iterative bi-directional training between the BSs and the UEs to carry out the alternating optimization of the precoding and combining strategies. More specifically, each bi-directional training iteration of the distributed precoding design via backhaul signaling described in Section~\ref{sec:PROB_distr} involves one uplink signaling resource and one downlink signaling resource (i.e., DL and UL-1 in steps S.1 and S.3, respectively, of Algorithm~\ref{alg:distr_bh}). On the other hand, the proposed distributed precoding design via OTA signaling introduces an additional uplink signaling resource at each bi-directional training iteration (i.e., UL-2 in step S.4 of Algorithm~\ref{alg:distr_ota}). This allows each BS to acquire the missing cross-term information necessary for the distributed precoding design without any backhaul signaling for CSI exchange among the BSs. Building on~\cite{Tol19}, the proposed OTA signaling (consisting of DL, UL-1, and UL-2) can be easily integrated into the flexible 5G 3GPP NR frame/slot structure~\cite{3GPP_38.211} as described next; we refer to~\cite{Shi14,Kal18,Jay18} for more references on iterative bi-directional training.

\begin{figure*}[t!]
\centering
\includegraphics[scale=1]{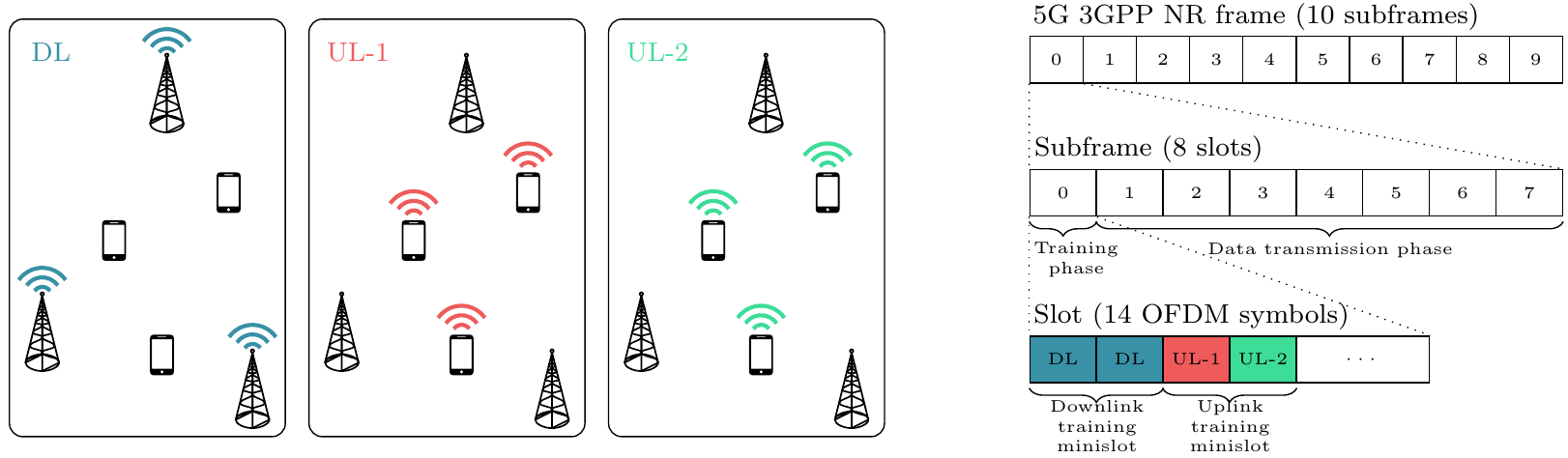}

\vspace{-0.5mm}

\hspace{1.9cm} \footnotesize{(a)} \hspace{8.3cm} \footnotesize{(b)} \vspace{-1mm}

\caption{(a) Schematic illustration of the proposed OTA signaling scheme; (b) example of how it can be integrated into the flexible 5G 3GPP NR frame/slot structure (right).} \label{fig:ota} \vspace{-7mm}
\end{figure*}

In the 5G 3GPP NR frame, each of the $10$~subframes consists of $8$~slots spanning $14$~orthogonal frequency division multiplexing (OFDM) symbols and can be conveniently divided into training phase and data transmission phase~\cite{Tol19}. In this regard, the 5G 3GPP NR standard defines the minislot structure with duration of minimum $2$~OFDM symbols, which can be flexibly constructed to accommodate either uplink or downlink training. Hence, each training minislot can contain two downlink signaling resources (i.e., DL twice) or two uplink signaling resources (i.e., UL-1 twice, UL-2 twice, or UL-1 and UL-2).\footnote{To fit each (uplink or downlink) signaling resource into one OFDM symbol, the pilot length $\tau$ must be less than the number of available subcarriers.} In the example in Figure~\ref{fig:ota}, the training phase of Algorithm~\ref{alg:distr_ota} takes place during the first slot, whereas the remaining $7$~slots are dedicated to the data transmission. Here, the training slot is constructed as a sequence of downlink and uplink minislots, with each uplink minislot including UL-1 and UL-2; by doing so, one training slot can contain up to $3$~bi-directional training iterations, which have a total duration of $12$~OFDM symbols.\footnote{The switching time between uplink and downlink signaling should be also taken into account. In this example, the $2$~remaining OFDM symbols can be conveniently used to separate downlink and uplink training minislots.} In general, the amount of OTA signaling and its placement within the 5G 3GPP NR frame can be adjusted based on rate and latency requirements. For example, for enhanced mobile broadband services, one can allow an extended training phase consisting of multiple slots to accommodate a large number of iterations followed by a long data transmission phase. On the other hand, if the latency is the primary requirement, it is more desirable to alternate short training phases (e.g., of only one slot) and brief data transmission phases.

Now, let us briefly compare Algorithms~\ref{alg:distr_ota}~and~\ref{alg:distr_bh} in terms of signaling overhead. We begin by considering that, under the 5G 3GPP NR frame/slot structure, each bi-directional training iteration of the distributed precoding design via backhaul signaling contains one uplink training minislot  including UL-1 twice. If each bi-directional training iteration of Algorithm~\ref{alg:distr_ota} contains one uplink training minislot including UL-1 and UL-2 (as in the example in Figure~\ref{fig:ota}), then there is no extra signaling overhead as compared with Algorithm~\ref{alg:distr_bh}; however, a small performance loss is expected since, in the latter, the uplink pilot-aided channel estimation is more accurate. On the other hand, if each bi-directional training iteration of Algorithm~\ref{alg:distr_ota} contains two uplink training minislots including UL-1 twice and UL-2 twice, respectively, then there is a $50\%$~increase in the signaling overhead with respect to Algorithm~\ref{alg:distr_bh} with no loss in estimation accuracy.

\subsection{UE Transmit Power Scaling} \label{sec:OTA_power}

During the uplink training, each UE~$k$ uses its combining vector $\v_{k}$ as precoder in UL-1 and UL-2 (see \eqref{eq:X_k_ul1} and \eqref{eq:X_k_ul2}, respectively). However, the power of $\v_{k}$ computed as in \eqref{eq:v_k_distr_imp} can be quite significant as it is roughly inversely proportional to the received signal power during the downlink pilot-aided channel estimation. Hence, the power scaling factors $\beta^{\ulA}$ and $\beta^{\ulB}$ (equal for all the UEs) in \eqref{eq:X_k_ul1} and \eqref{eq:X_k_ul2}, respectively, must be chosen to ensure that the uplink training complies with the UE transmit power constraint. In fact, without proper UE transmit power scaling, $\X_{k}^{\ulA}$ and $\X_{k}^{\ulB}$ in \eqref{eq:X_k_ul1} and \eqref{eq:X_k_ul2}, respectively, will most likely exceed the maximum transmit power $\rho_{\ue}$, unless UE~$k$ is located in close proximity of a BS. Finally, upon receiving $\Y_{b}^{\ulA}$ and $\Y_{b}^{\ulB}$ in \eqref{eq:Y_b_ul1} and \eqref{eq:Y_b_ul2}, respectively, each BS~$b$ scales back the receive signals to obtain the desired CSI, which results in an amplification of the AWGN terms in \eqref{eq:h_bk_hat} and \eqref{eq:n_bk_ul2}, respectively.

The power scaling factors $\beta^{\ulA}$ and $\beta^{\ulB}$ can be determined by the BSs or the CPU and transmitted to the UEs via suitable feedback channels. Note that adopting the same power scaling factors for all the UEs is crucial to keep the interdependencies among the UE channels intact, although this may result in some UEs transmitting with power much lower than $\rho_{\ue}$ during the uplink training. More specifically, the power scaling factors can be obtained based on statistical information such as the average received signal power of the UEs across the network (which in turn depends on the number, the placement, and the transmit power of the BSs), as done in our numerical results in Section~\ref{sec:NUM}.

\section{Numerical Results and Discussion} \label{sec:NUM}

In this section, we present numerical results to compare the performance of the proposed distributed precoding design via OTA signaling in Algorithm~\ref{alg:distr_ota} (Distributed--OTA) with: \textit{i)}~the local MMSE precoding (Local MMSE), \textit{ii)}~the centralized precoding design in Algorithm~\ref{alg:centr} (Centralized), \textit{iii)}~the centralized precoding design with iterative bi-directional training in Algorithm~\ref{alg:centr_it} (Centralized--iterative), and \textit{iv)}~the distributed precoding design via backhaul signaling in Algorithm~\ref{alg:distr_bh} (Distributed--backhaul). For the latter, we assume~that the backhaul signaling introduces a delay of only one bi-directional training iteration in the CSI exchange among the BSs; furthermore, we assume perfect backhaul links and, thus, no quantization errors in the backhaul signaling (the impact of this factor is evaluated in~\cite{Kal18}). Note that both assumptions are quite optimistic and favor the Distributed--backhaul over the proposed Distributed--OTA.

The simulation setup consists of $B = 25$ BSs equipped with $M = 4$ antennas (unless otherwise stated) placed on a square grid with inter-site distance of $100$~m and height of $10$~m. The BSs serve $K = 16$ UEs (unless otherwise stated) equipped with $N = 2$ antennas, which are randomly dropped in the same area. As in~\cite{Gou20,Atz20}, the channel model includes i.i.d. Rayleigh fading and power-law pathloss with each channel generated as $\mathrm{vec}(\H_{b,k}) \sim \setC \setN (0, \delta_{b,k} \I_{MN})$, where $\delta_{b,k} \ [\textnormal{dB}] \triangleq -30.5 -36.7 \log_{10} (r_{b,k})$ and where $r_{b,k}$ is the distance between BS~$b$ and UE~$k$. The transmit powers at the BSs and the UEs are fixed to $\rho_{\bs} = 30$~dBm and $\rho_{\ue} = 20$~dBm, respectively, whereas the AWGN powers at the BSs and at the UEs are fixed to $\{\sigma_{b}^{2} = -95~\textrm{dBm}\}_{b \in \setB}$ and $\{\sigma_{k}^{2} = -95~\textrm{dBm}\}_{k \in \setK}$, respectively (unless otherwise stated). As performance metric, we evaluate the average sum rate $\Exp[R]$ obtained via Monte~Carlo simulations of the sum rate in \eqref{eq:R} with $10^3$ independent UE drops. As discussed in Section~\ref{sec:SM}, this average sum rate represents an upper bound on the system performance, which is achievable if perfect global CSI is available at all the BSs and which becomes increasingly tight as the duration of the coherence block increases~\cite{Cai18}. Observe that replacing this upper bound with any ergodic achievable rate expression (such as the one given in~\cite[Lem.~3]{Cai18} would have the same impact on all the algorithms and, therefore, the relative performance gaps between the different precoding schemes would not be affected.

\begin{figure}[t!]
\begin{minipage}[c]{0.49\textwidth}
\centering
\vspace{-1.2mm} \includegraphics[scale=0.8]{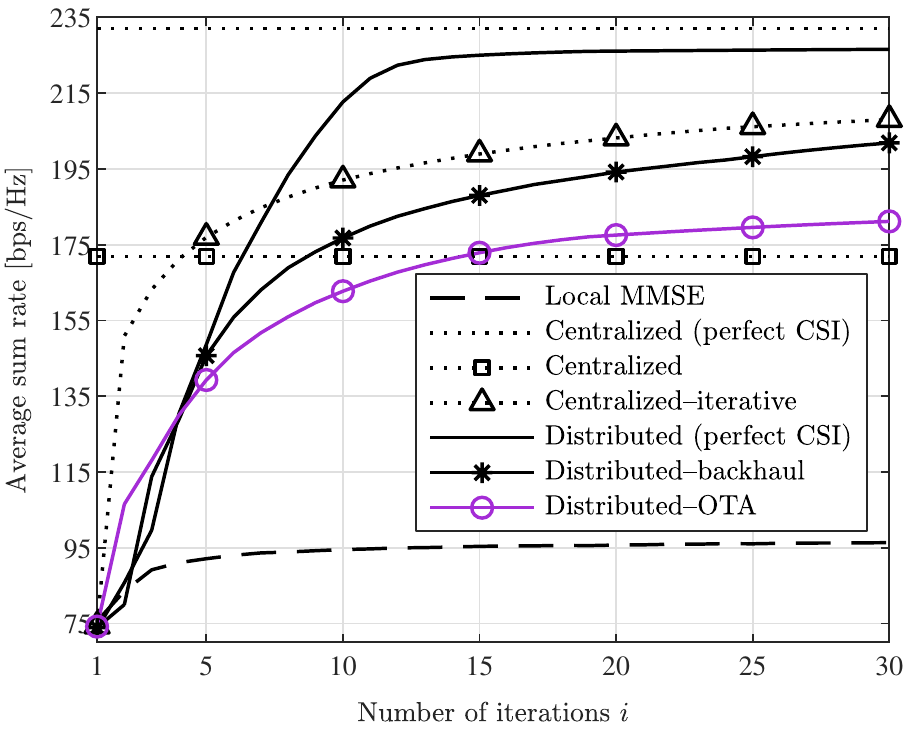}
\caption{Average sum rate versus number of bi-directional training iterations with orthogonal pilots.} \label{fig:rate_vs_i}
\end{minipage}
\hspace{1mm}
\begin{minipage}[c]{0.49\textwidth}
\centering
\vspace{-1.4mm}
\includegraphics[scale=0.8]{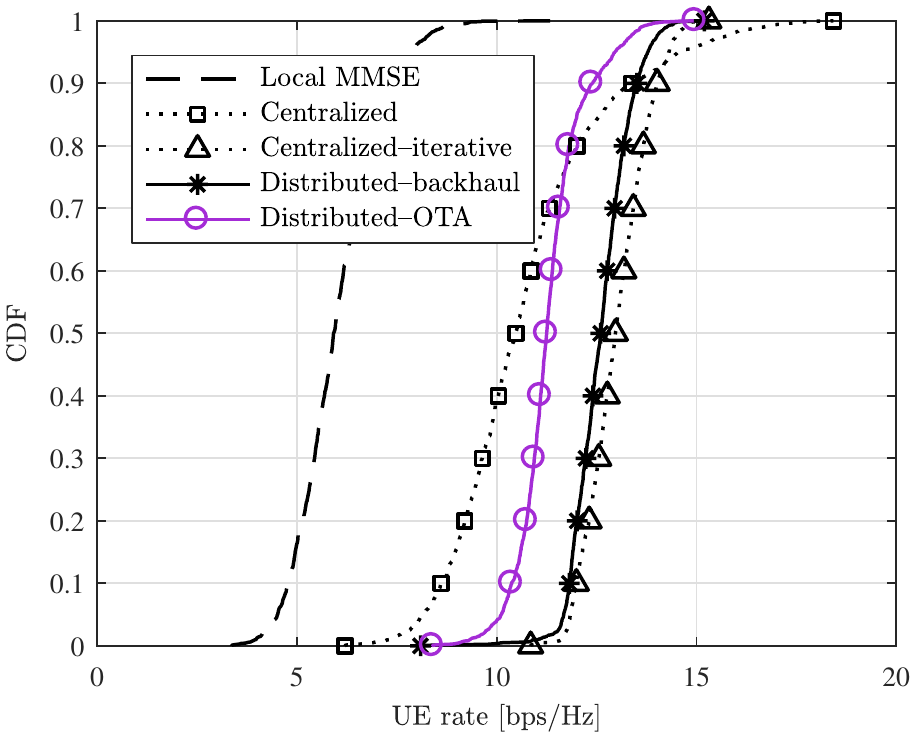}
\caption{CDF of the per-UE rate after convergence with orthogonal pilots.} \label{fig:cdf_vs_rate}
\end{minipage} \vspace{-7mm}
\end{figure}

We begin by considering the case of orthogonal pilots. Figure~\ref{fig:rate_vs_i} illustrates the average sum rate (without considering the signaling overhead) versus the number of bi-directional training iterations; here, the centralized and the distributed precoding designs with perfect CSI are also included for comparative purposes. The proposed Distributed--OTA achieves a performance increase with respect to the Local MMSE of about $55\%$ after just $5$ iterations and of about $90\%$ after convergence. Furthermore, it reaches the performance of the Centralized within $14$ iterations. As discussed in Remark~\ref{rem:averaging}, under imperfect CSI, the distributed precoding designs can outperform the Centralized, which relies on a single pilot-aided CSI acquisition: under the assumption of orthogonal pilots, this performance gain arises from the noise averaging associated with the bi-directional training. In this regard, we observe that the gap between the Centralized (perfect CSI) and the Centralized is substantially larger than in the distributed schemes based on bi-directional training, which confirms the advantage brought by the distributed precoding design under imperfect CSI. Let us now compare the proposed Distributed--OTA with the Distributed--backhaul. The performance loss of the former with respect to the latter stems from the OTA acquisition of the cross-term information, which combines three noisy signaling resources (namely, DL, UL-1, and UL-2). However, during the first few iterations, the Distributed--OTA converges faster than the Distributed--backhaul, which suffers from the delayed backhaul update. In this regard, the average sum rate achieved by the two schemes after $5$ iterations is nearly the same. Above all, eliminating the need for backhaul signaling for CSI exchange brings huge practical benefits that justify this slight performance degradation; otherwise, this gap can be also bridged by means of power-boosted uplink signaling. As expected, the Centralized--iterative produces the best average sum rate under imperfect CSI; however, we remark that this scheme is highly impractical due to the burdensome backhaul signaling between the BSs and the CPU. Figure~\ref{fig:cdf_vs_rate} shows the cumulative distribution function (CDF) of the per-UE rates. It is easy to observe that the distributed precoding design treats the UEs more fairly as compared with the Centralized. In particular, each UE is served with a rate higher than $10$~bps/Hz with probability $0.98$ for the Distributed--OTA and with probability $0.6$ for the Centralized.

\begin{figure}[t!]
\centering
\includegraphics[scale=0.8]{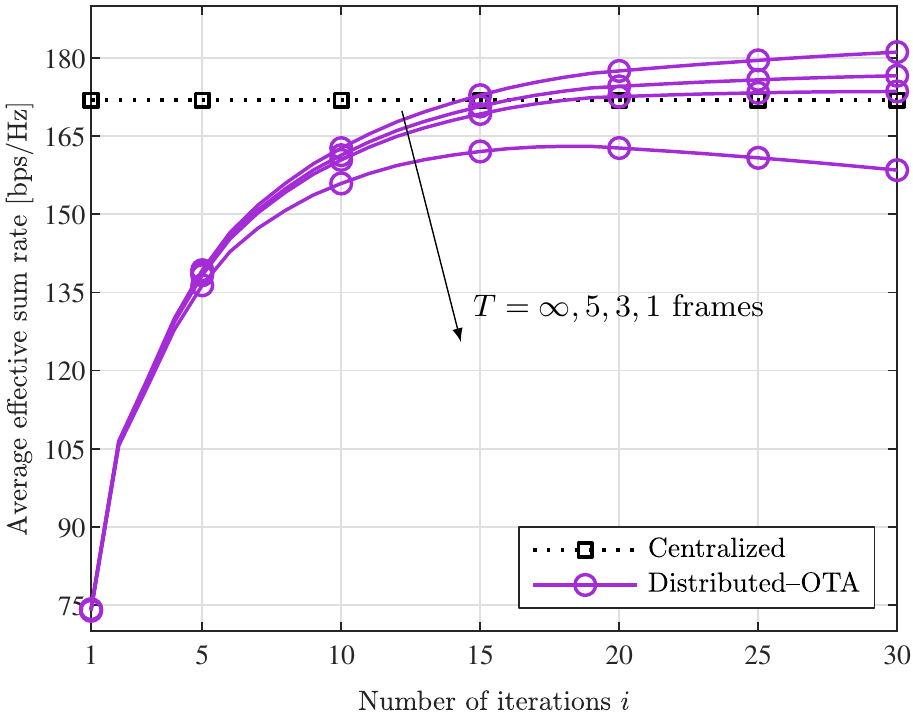}
\caption{Average effective sum rate versus number of bi-directional training iterations with orthogonal pilots and scheduling blocks of different duration (where $T=\infty$ corresponds to the case of no signaling overhead).} \label{fig:rate_vs_i_overhead} \vspace{-7mm}
\end{figure}

Now, let us evaluate the impact of the signaling overhead. Under the 5G 3GPP NR frame/slot structure, we consider scheduling blocks with duration of $T$~frames (where each frame has~duration of $1120$~OFDM symbols), during which the channels remain fixed. The whole training phase occurs at the beginning of the scheduling block and each bi-directional training iteration occupies $\frac{14}{3} \simeq 4.67$~OFDM symbols (as detailed in Section~\ref{sec:OTA_frame}). Hence, the effective sum rate resulting from taking into account the signaling overhead after $i$~bi-directional training iterations is given~by
\begin{align}
R_{\mathrm{eff}}^{(i)} \triangleq \bigg( 1 - \frac{4.67 i}{1120 T} \bigg) R^{(i)}
\end{align}
with $R^{(i)}$ being the sum rate after $i$~iterations (obtained as in \eqref{eq:R}). Figure~\ref{fig:rate_vs_i_overhead} plots the average effective sum rate of the proposed Distributed--OTA versus the number of bi-directional training iterations. The performance loss due to the signaling overhead is very modest for $T=5$~frames, and the Centralized can be still outperformed for $T=3$~frames. On the other hand, for $T=1$~frames, $i=19$ is the optimal number of bi-directional training iterations after which the average effective sum rate starts decreasing. It is worth noting that the number of bi-directional training iterations (and, thus, the overall signaling overhead) would be greatly reduced in case of time-correlated channels with semi-persistent UE scheduling. In fact, assuming that the UE scheduling remains essentially unchanged between scheduling blocks, the precoding and combining vectors need not be computed from scratch in each scheduling block and a brand new training phase is required only when the UE scheduling changes; we refer to~\cite{Kal18,Tol19} for more details.

\begin{figure}[t!]
\centering
\includegraphics[scale=0.8]{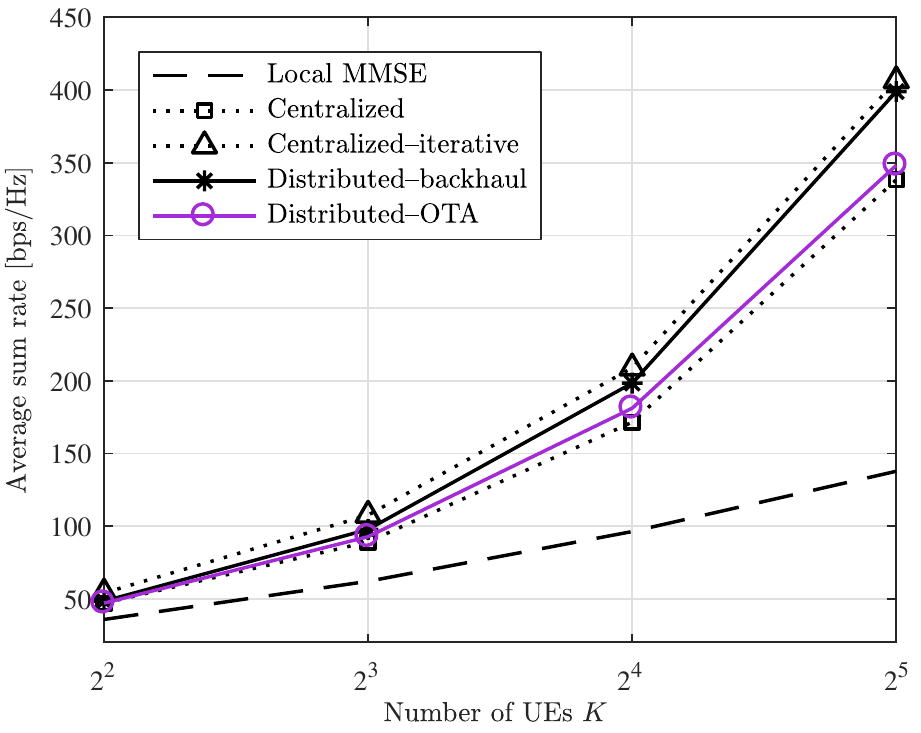} \hspace{5mm}
\includegraphics[scale=0.8]{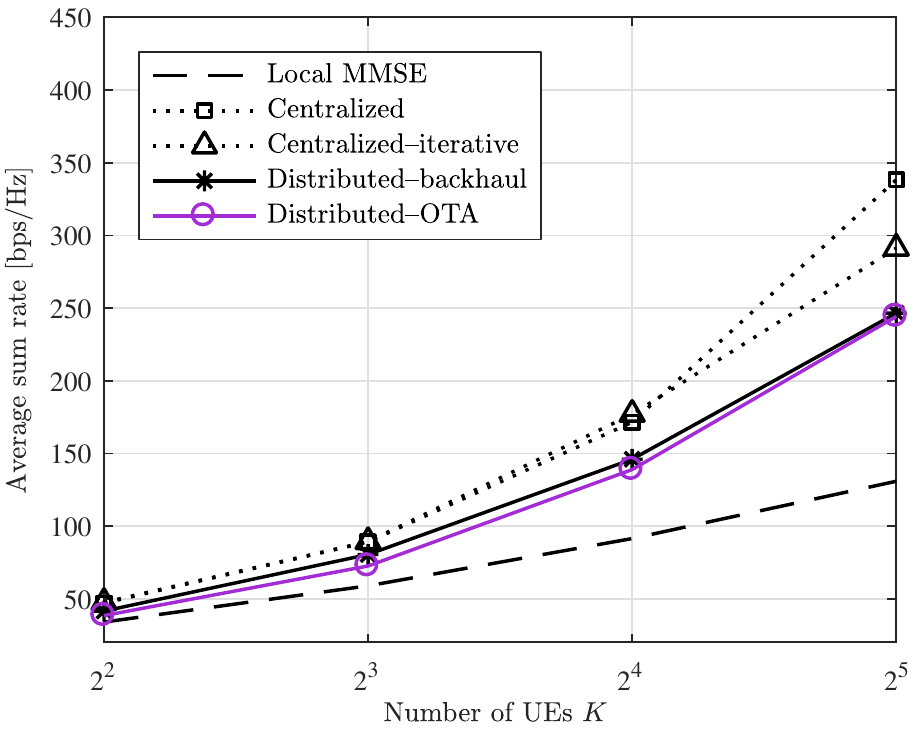}

\footnotesize{(a)} \hspace{8cm} \footnotesize{(b)} \vspace{-1mm}

\caption{Average sum rate versus number of UEs with orthogonal pilots: (a) after convergence; (b) after $5$ bi-directional training iterations.} \label{fig:rate_vs_K} \vspace{-7mm}
\end{figure}

\begin{figure}[t!]
\begin{minipage}[c]{0.49\textwidth}
\centering
\vspace{-1.2mm} \includegraphics[scale=0.8]{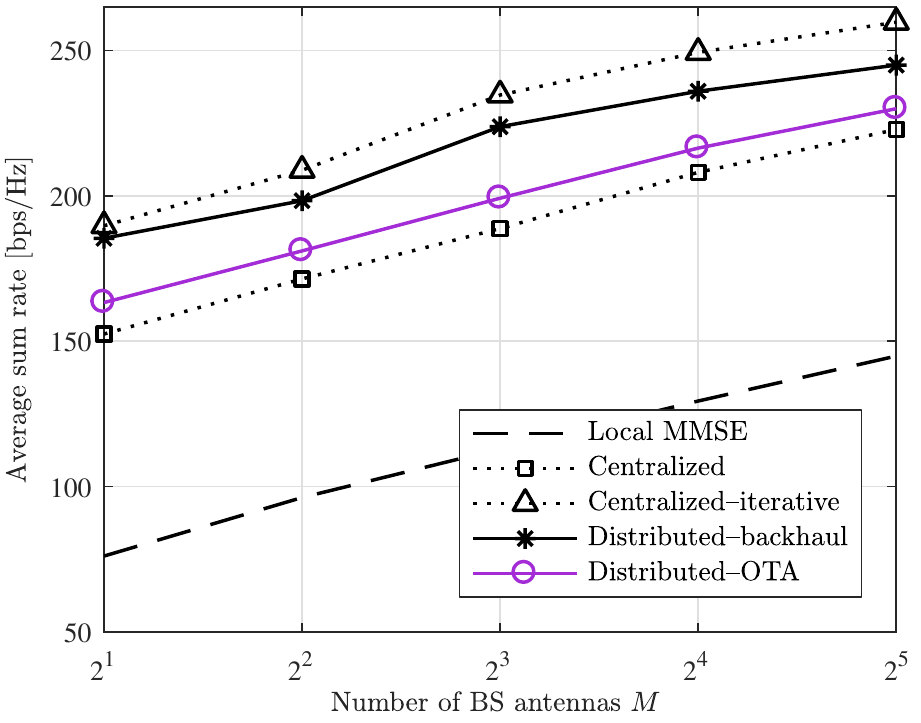}
\caption{Average sum rate versus number of antennas at each BS.} \label{fig:rate_vs_M}
\end{minipage}
\hspace{1mm}
\begin{minipage}[c]{0.49\textwidth}
\centering
\vspace{-1mm}
\includegraphics[scale=0.8]{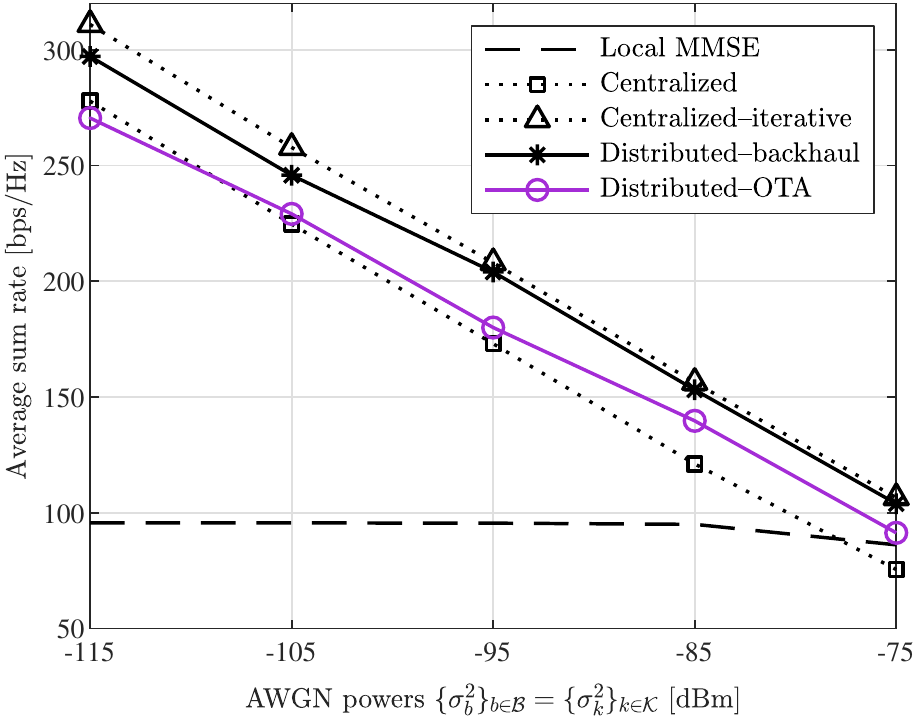}
\caption{Average sum rate versus AWGN powers at the BSs and at the UEs.} \label{fig:rate_vs_noise}
\end{minipage} \vspace{-7mm}
\end{figure}

Figure~\ref{fig:rate_vs_K} illustrates the average sum rate versus the number of UEs~$K$. First of all, in Figure~\ref{fig:rate_vs_K}(a), we observe that the performance gain brought by cooperative precoding design over the Local MMSE increases significantly with the spatial load. Besides, the proposed Distributed--OTA after convergence outperforms the Centralized for any value of $K$. On the other hand, in Figure~\ref{fig:rate_vs_K}(b), the performance gap between the Distributed--OTA and the Distributed--backhaul after $5$~iterations is remarkably small. In this regard, the average sum rate obtained with the Distributed--OTA is about $90\%$ higher than that obtained with the Local MMSE. Figure~\ref{fig:rate_vs_M} depicts the average sum rate versus the number of BS antennas~$M$. As $M$ increases, each BS has more degrees of freedom to tackle the interference locally and, hence, the performance gain brought by cooperative precoding design over the Local MMSE remains approximately constant. However, note that cell-free massive MIMO generally assumes a low-to-moderate number of antennas at the BSs~\cite{Zha19}, for which cooperative precoding design is highly beneficial. It is worth remarking that, in Figures~\ref{fig:rate_vs_K}~and~\ref{fig:rate_vs_M}, the amount of signaling scales with the number of UEs~$K$ and the number of BSs~$B$, respectively, for all the schemes except for the proposed Distributed--OTA, as the latter does not involve any CSI exchange via backhaul signaling. Figure~\ref{fig:rate_vs_noise} plots the average sum rate versus the AWGN powers at the BSs and at the UEs, which are the same for the channel estimation phase and for the data transmission phase. The performance of the Local MMSE is roughly constant over the whole range and is comparable to that of the other more complex schemes only for very low signal-to-noise ratio (SNR). On the other hand, the average sum rate obtained with both centralized and distributed precoding design increases considerably with the SNR. Remarkably, the Distributed--OTA outperforms the Centralized for AWGN powers as low as $-110$~dBm, below which the latter benefits from the very accurate channel estimation.

\begin{figure}[t!]
\centering
\includegraphics[scale=0.8]{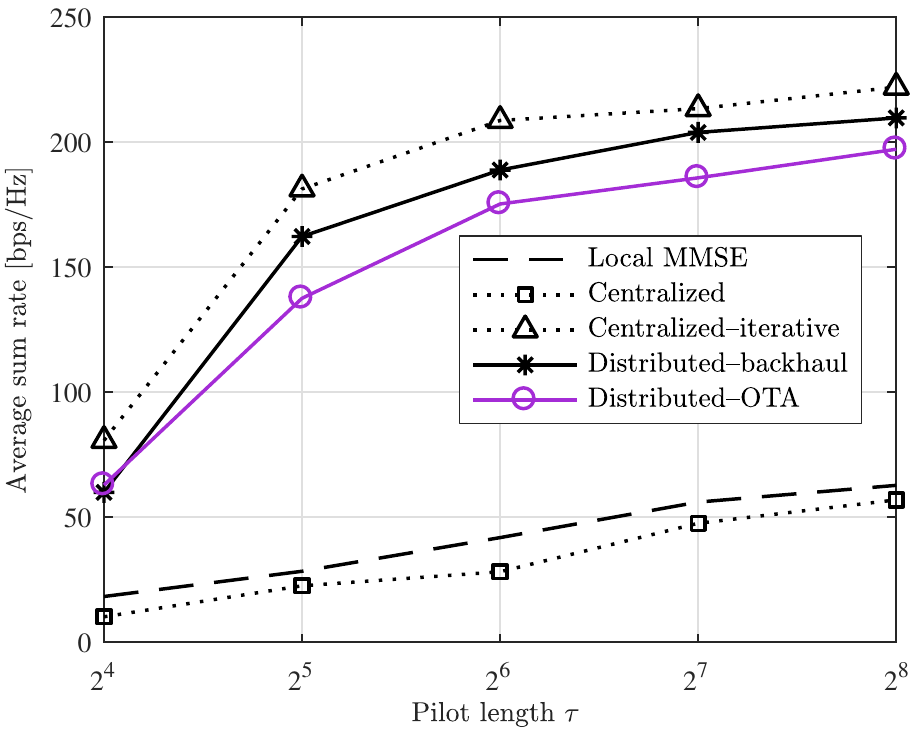}
\caption{Average sum rate versus pilot length with non-orthogonal random pilots.} \label{fig:rate_vs_tau} \vspace{-7mm}
\end{figure}

Lastly, we consider a pilot-contaminated scenario by assuming non-orthogonal random pilots. Figure~\ref{fig:rate_vs_tau} plots the average sum rate versus the pilot length~$\tau$; here, we still impose $\{\P_{k}^{\herm} \P_{k} = \tau \I_{N}\}_{k \in \setK}$ for the Centralized, i.e., the antenna-specific pilots within each UE~$k$ are orthogonal. First of all, the Centralized is extremely sensitive to the pilot contamination in \eqref{eq:H_bk_hat} as it relies on a single pilot-aided CSI acquisition: for this reason, it performs very poorly (even worse than the Local MMSE). On the other hand, in the distributed schemes, the precoding and combining vectors are directly estimated at each bi-directional training iteration, as detailed in Remark~\ref{rem:averaging} (see also~\cite{Kal18}). This provides greatly improved robustness against pilot contamination and the ideal performance can be approached by increasing the pilot length. Note that the use of non-orthogonal random pilots does not require any centralized coordination and may be practical in large networks even if the pilot length allows to obtain orthogonal pilots for all the UEs.

\section{Conclusions} \label{sec:CONCL}

In this paper, we proposed the first distributed framework for cooperative precoding design in cell-free massive MIMO (and, more generally, in JT-CoMP) systems that entirely eliminates the need for backhaul signaling for CSI exchange. To do so, we presented a novel OTA signaling mechanism that allows each BS to obtain the same cross-term information that is traditionally exchanged among the BSs via backhaul signaling. This was achieved by introducing a new uplink signaling resource and a new CSI combining mechanism that complement the existing uplink and downlink pilot-aided channel estimations. Remarkably, the amount of OTA signaling does not grow with the number of BSs or UEs, which makes our distributed precoding design scalable to any network setup. In addition, the proposed OTA signaling does not introduce any delay in the CSI exchange among the BSs and can be easily integrated into the flexible 5G 3GPP NR frame/slot structure. Note that, although this paper targeted the weighted sum MSE minimization problem, the proposed OTA signaling mechanism can be applied to any network optimization utility. Our numerical results showed fast convergence and significant performance gains over non-cooperative precoding design; the proposed scheme can also outperform the centralized precoding design due to its robustness against both pilot contamination and noisy channel estimation. In conclusion, by eliminating the need for backhaul signaling for CSI exchange, our contribution aims at facilitating the practical implementation of cell-free massive MIMO and JT-CoMP in future 5G and beyond-5G systems.

\appendices

\section{Equivalence Between the Centralized and the Distributed Precoding Solutions} \label{app:equiv}

Assuming the simple case of $B=2$ BSs, we can rewrite \eqref{eq:w_bk} for $b=1$ as
\begin{align}
\w_{1,k} & = (\Phib_{1 1} + \lambda_{1} \I_{M})^{-1} (\omega_{k} \h_{1,k} - \Phib_{12} \w_{2,k}) \\
& = \omega_{k} \big( \Phib_{11} + \lambda_{1} \I_{M} - \Phib_{12} (\Phib_{22} + \lambda_{2} \I_{M})^{-1} \Phib_{12}^{\herm} \big)^{-1} \big( \h_{1,k} - \Phib_{12} (\Phib_{22} + \lambda_{2} \I_{M})^{-1} \h_{2,k} \big)
\end{align}
The equivalence $\w_{k} = [\w_{1,k}^{\tran}, \w_{2,k}^{\tran}]^{\tran}, \forall k \in \setK$, i.e.,
\begin{align} \label{eq:equiv}
\begin{bmatrix}
\Phib_{1 1} + \lambda_{1} \I_{M}	& \Phib_{1 2} \\
\Phib_{1 2}^{\herm}          		& \Phib_{2 2} + \lambda_{2} \I_{M}
\end{bmatrix}^{-1} \begin{bmatrix}
\h_{1,k} \\
\h_{2,k}
\end{bmatrix} \qquad \qquad \qquad \qquad \qquad \qquad \qquad \qquad \qquad \qquad \nonumber  \\  
= \begin{bmatrix}
\big( \Phib_{1 1} + \lambda_{1} \I_{M} - \Phib_{1 2} (\Phib_{2 2} + \lambda_{2} \I_{M})^{-1} \Phib_{1 2}^{\herm} \big)^{-1} \big( \h_{1,k} - \Phib_{1 2} (\Phib_{2 2} + \lambda_{2} \I_{M})^{-1} \h_{2,k} \big) \\
\big( \Phib_{2 2} + \lambda_{2} \I_{M} - \Phib_{2 1} (\Phib_{1 1} + \lambda_{1} \I_{M})^{-1} \Phib_{2 1}^{\herm} \big)^{-1} \big( \h_{2,k} - \Phib_{2 1} (\Phib_{1 1} + \lambda_{1} \I_{M})^{-1} \h_{1,k} \big)
\end{bmatrix}
\end{align}
follows from applying the Schur complement to the inverse matrix on the left-hand side of \eqref{eq:equiv}.

\bibliographystyle{IEEEtran}
\bibliography{IEEEabrv,refs}

\end{document}

%% file: Definitions.tex
\newcommand{\g}{\mathbf{g}}
\newcommand{\h}{\mathbf{h}}

\newcommand{\n}{\mathbf{n}}

\newcommand{\p}{\mathbf{p}}

\renewcommand{\v}{\mathbf{v}}
\newcommand{\w}{\mathbf{w}}
\newcommand{\x}{\mathbf{x}}
\newcommand{\y}{\mathbf{y}}
\newcommand{\z}{\mathbf{z}}

\newcommand{\0}{\mathbf{0}}

\newcommand{\E}{\mathbf{E}}

\renewcommand{\H}{\mathbf{H}}
\newcommand{\I}{\mathbf{I}}

\newcommand{\N}{\mathbf{N}}

\renewcommand{\P}{\mathbf{P}}

\newcommand{\W}{\mathbf{W}}
\newcommand{\X}{\mathbf{X}}
\newcommand{\Y}{\mathbf{Y}}
\newcommand{\Z}{\mathbf{Z}}

\newcommand{\xib}{\boldsymbol{\xi}}

\newcommand{\Phib}{\mathbf{\Phi}}
\newcommand{\Psib}{\mathbf{\Psi}}
\newcommand{\Omegab}{\mathbf{\Omega}}

\newcommand{\setB}{\mathcal{B}}
\newcommand{\setC}{\mathcal{C}}

\newcommand{\setK}{\mathcal{K}}
\newcommand{\setL}{\mathcal{L}}

\newcommand{\setN}{\mathcal{N}}
\newcommand{\setO}{\mathcal{O}}

\newcommand{\Real}{\mbox{$\mathbb{R}$}}
\newcommand{\Compl}{\mbox{$\mathbb{C}$}}
\newcommand{\rmF}{\mathrm{F}}

\newcommand{\Diag}{\mathrm{Diag}}

\newcommand{\Exp}{\mathbb{E}}
\newcommand{\herm}{\mathrm{H}}

\renewcommand{\Re}{\mathrm{Re}}
\newcommand{\tr}{\mathrm{tr}}
\newcommand{\tran}{\mathrm{T}}